\documentclass[10pt,twocolumn]{article}
\usepackage{amsmath}
\usepackage{amsfonts}
\usepackage{amssymb}
\usepackage{color}
\usepackage{xcolor}
\usepackage{colortbl}
\RequirePackage{pgfplots}
\usepackage{tikz}
\usepackage{tabularx}
\newcolumntype{Y}{>{\centering\arraybackslash}X}
\newcommand\setrow[1]{\gdef\rowmac{#1}#1\ignorespaces}
\newcommand\clearrow{\global\let\rowmac\relax}\clearrow
\usepackage[letterspace]{microtype}
\usepackage{booktabs}
\usepackage[notes, notetype=endonly,backend=bibtex,autocite=footnote]{biblatex-chicago}  
\usepackage{enotez,refcount} 
\setenotez{backref=true}
\usepackage{needspace}

\usepackage{enumitem}
\newlist{aublistforenotez}{itemize}{1}
\setlist[aublistforenotez]{parsep=-\parskip, itemsep=-\parskip, topsep=0pt,itemindent=\dimexpr\labelwidth\relax,leftmargin=0pt}
\DeclareInstance{enotez-list}{aubcustomlist}{list}
{
heading =\sparagraph{\togglefont #1},
format = \footnotesize,
number-format  = \normalfont,
number =  \enmark{#1},
number = \textsuperscript{#1},
list-type = aublistforenotez
}

\renewcommand{\autocite}[2][]{\endnote{\cite[#1]{#2}}}
\usepackage{booktabs}
\usepackage{anyfontsize}
\usepackage[hidelinks]{hyperref}
\hypersetup{pdftitle={A Critical Review of the Baseline Soldier Physical Readiness Requirements Study}, pdfauthor={Kyle Novak}, pdfkeywords={}, pdfproducer={},
 pdfcreator={}}
\usepackage{newtxtext}
\usepackage{newtxmath}
\usepackage[T1]{fontenc}
\usepackage{gillius2}
\usepackage{helvet}
\usepackage{tgadventor} 
\def\togglefont{\fontfamily{phv}\selectfont}
\def\togglefont{\fontfamily{qag}\selectfont}

\usepackage{array}
\newcolumntype{$}{>{\global\let\currentrowstyle\relax}}
\newcolumntype{^}{>{\currentrowstyle}}

\usepackage{time}
\usepackage{fancyhdr}
\newlength{\bbarsep}
\newcommand\bbar[5]{
  \addplot[line width=1.5mm,line cap=round] coordinates {(#3,#1) (#4,#1)};
  \addplot[line width=1.5mm,color=white] coordinates {(#5+0.25,#1) (#5-0.25,#1)};

  \node[align=right,left=0.25em] at  (axis cs:0,#1) {#2};
}
\definecolor{newgray}{rgb}{0,0,0}

	\newcommand{\Var}[1]{\mathrm{Var}{\left[#1\right]}}
	\newcommand{\Cov}[1]{\mathrm{Cov}{\left[#1\right]}}

\usepackage{setspace}
\usepackage{titlesec}

\def\WTST{WTST}
\definecolor{Blue}{rgb}{0.2,0.2,.5}
\definecolor{FontBlue}{rgb}{0.3,0.3,.8}
\titleformat{\section}{\centering\togglefont\bfseries\color{FontBlue}}{\thesection.}{\enspace}{}{}

\def\NA{$*$}
\newcommand\textQT[1]{``#1''}

\newcommand\sparagraph[1]{\paragraph{\color{FontBlue}\togglefont#1.}}

\begin{document}
\title{\vspace{-0.5\baselineskip}\fontsize{20}{24}\togglefont{\color{FontBlue}A Critical Review of the Baseline Soldier\\ Physical Readiness Requirements Study\\[-0.7\baselineskip]}}
\date{}

\twocolumn[
  \begin{@twocolumnfalse}
  \maketitle
  \end{@twocolumnfalse}
]
\thispagestyle{fancy}

\sparagraph{Executive Summary}
The Army's Baseline Soldier Physical Readiness Requirements Study (BSPRRS) was a multiyear effort designed to inform evidence-based change in the Army's physical fitness test of record from a gender- and age-specific standard to a gender- and age-neutral standard based on recurring, physically demanding  tasks encountered by soldiers. The study determined Warrior Tasks and Battle Drills and Common Soldier Tasks (WTBD/CST); developed a set of four vignettes that simulated those tasks; identified physical fitness test events that were modeled to predict performance on these vignettes; and finally sought to validate that those test events were indeed predictive. The physical fitness test events determined by BSPRRS became the six-event Army Combat Fitness Test (ACFT). While the Army may have been well-intentioned in its desire to develop an evidence-based,  operationally predictive, and gender-neutral fitness standard, close examination of the BSPRRS model reveals  critical mistakes and the lack of rigorous cross-validation.
Using the ACFT and the BSPRRS model to guide decisions about soldiers' physical fitness may contrarily make for a less lethal Army. 

The 2015 NDAA requires that gender-neutral occupational standards ``accurately predict performance of actual, regular, and recurring duties of a military occupation'' and are ``applied equitably to measure individual capabilities.''   While the Army often claims that the ACFT is over 80 percent predictive,  this claim is a misrepresentation of the flawed BSPRRS model.  In fact, because researchers likely compared ACFT events against only one of the four WTBD/CST simulation vignettes,  only two of the six ACFT test events were modeled as  significant predictors of WTBD/CST performance. And, in demonstrating  predictive accuracy of the ACFT, the Army used only 136 male soldiers and a mere 16 female soldiers, all volunteers with an average age of 24 years, to represent the entire Army. The Army's new goal of collecting two million data points on soldiers taking ACFT without first addressing the errors in the BSPRRS model or validating  the predictive performance of the ACFT is simply shooting in the dark. 

Furthermore,  the ACFT is  not applied equitably.
In initial trials with over 14 thousand soldiers, sixty-five percent of all women failed the ACFT, primarily because of the leg tuck test event, compared to   ten percent of male soldiers. But, according to data from the Army's own study, leg tucks are  not predictive at all of actual, regular, and recurring duties. Indeed, using leg tucks as a criterion creates an unfair adverse impact. 
The University of Iowa Virtual Soldier Center, who peer reviewed BSPRRS,  criticized the Army for the lack of female participants in the study stating that due to the ``inherently unbalanced study design...determination of which tasks best predict or represent WTBD/CST performance could be influenced towards strategies used predominantly by men.'' 

Moreover, the ACFT may undermine military readiness. In moving from the original Army Physical Fitness Test (APFT) to the ACFT, the Army has made their fitness test of record 20 times easier for young male recruits and 1.3 times easier for young female recruits.  At the same time, the test is more difficult for older female soldiers, precisely those who are already disproportionately underrepresented in senior leadership positions. 

Finally, one must question whether the ACFT needs to be gender- and age-neutral at all. While occupational standards are required by the 2015 NDAA to be gender-neutral,  the Army emphasizes the role of the ACFT in Holistic Health and Fitness to combat obesity and reduce muscular-skeletal injuries. RAND and others have argued that fitness assessments meant ``to maintain a culture of military discipline, bearing, and appearance; to keep health care costs to a minimum; to ensure personnel are not likely to be hampered by chronic illness'' can and should be gender- and age-specific assessments.

\sparagraph{Background}
After forty years of testing  physical fitness using push-ups, sit-ups, and two-mile runs, in June 2020 the Army instituted a new assessment, called  the Army Combat Fitness Test (ACFT), to be the physical fitness test of record.\autocite{AD2020-06}  The ACFT consists of six test events: deadlift, standing power throw, push-ups, sprint-drag-carry, leg tucks, and the two-mile run. 
The Army is committed to the new test and is already spending  over \$78 million on the specialized equipment needed to administer it.%
\endnote{\href{https://www.army.mil/acft/\#faq-section-5}{https://www.army.mil/acft/\#faq-section-5}}
Every soldier regardless of age or gender is required to complete each test event to one of three minimum standards to pass the ACFT. This is a marked change from the previous Army Physical Fitness Test (APFT) that took age and gender into consideration.
Instead, the new standards are based on whether a soldier's military occupational specialty involves a  ``moderate,'' ``significant,'' or ``heavy'' amount of physical exertion.

In explaining the change, former Army Chief of Staff General Mark Milley said \textQT{This has everything to do with effectiveness in combat---that's why it's gender-neutral; that's why it's age-neutral. Combat is unforgiving. It doesn't matter how old you are. The enemy doesn't care. Before they shoot you, they don't say: `Hey, are you 25 or are you 45?' They don't do that. They just shoot you. And dead is dead. So we want to make sure that our soldiers are in top physical condition to withstand the rigors of ground combat.}\autocite{schogol_2018} 

During the first six months that the Army has been using the ACFT on approximately fourteen thousand soldiers,  sixty-five percent of  all female soldiers and ten percent of male soldiers have  failed to pass the ACFT at the minimum standard.%
\endnote{Gold-standard pass rates on the ACFT (from Army memo dated 09 July 2020 in response to Senator Gillibrand's request for information):\\
	\begin{tabularx}{1\linewidth}{lccccccc}\toprule
	 &FY19 & Oct & Nov & Dec & Jan & Feb & Average\\\midrule
	Female &21&37 & 29 & 33 & 34 & 40 & 32 \\
	Male  &  81& 91 & 89 & 91 & 90 & 90 & 89\\\midrule
	\end{tabularx}}
Army Secretary Mark Esper had stated earlier   \textQT{If you can't pass the Army combat-fitness test, then there's probably not a spot for you in the Army.}
General Milley had a similar sentiment:  \textQT{If you can't get in shape in 24 months, then maybe you should hit the road.}\autocite{cox_2018}

ACFT failures were largely due to failing to  do one leg tuck, the minimum number needed to pass that test event at the lowest, gold-standard level. Soldiers that must pass gray and black standards must complete even more leg tucks.
The Army has insisted that  \textQT{physiologically, there is no reason any healthy Soldier cannot perform a leg tuck, given 3--6 months of dedicated, regular and progressive training.} And  \textQT{achieving the ACFT GOLD standard takes some Soldiers about 3--4 months of focused training: everyone can meet this standard if properly trained and motivated. 
We are giving them $\sim$24 months.}
These statements run counter to advice by expert exercise physiologists.

General  Milley has praised the new ACFT stating that it has an \textQT{80 percent correlation to the physical activity that is expected of soldiers in the execution of ground combat.}  Other senior Army leaders promoting the new physical fitness test have made similar claims:  ``The APFT is a relatively poor predictor 
($\sim$40\%) 
of a Soldier's ability to execute high demand commonly occurring, critical Warrior Tasks and Battle Drills required of all Soldiers,'' ``The six-event ACFT has been scientifically validated through four years of extensive empirical research 
(R2 ACFT-High Demand CSTs $=$ 80\%) 
and is a better predictor of physical fitness associated with high physical demand common Soldier tasks,
and 
 ``Army Combat Fitness Test (ACFT) $\sim$80\% ability to predict WTBD/CST performance.''
These statements are simply not true.

\begin{table*}
\begin{tabularx}{1\linewidth}{cclX}\toprule
Phase& Year & Location & Purpose\\\midrule
1 & 2012--2013 & --- & Systematic literature review performed by U.S.~Army Public Health Center\\
2 & 2012--2013 & Multiple & Determine the physically demanding,
commonly occurring and critical Warrior Tasks and
Battle Drills and Common Soldier Tasks (WTBD/CST) \\
3 & 2013--2014 &  \parbox[t]{.12\linewidth}{Fort Carson} & Identify physical characteristics associated
with each WTBD/CST to develop a Warrior Task Simulation Test (\WTST)\\
4 & 2014--2015 & Fort Riley &  Determine which  fitness test events best predict \WTST\ performance\\
5  & 2015--2017 &  Fort Benning & Validate whether  fitness test events can accurately predict ability to execute WTBD/CST, are safe to perform, legally defensible, and acceptable\\\midrule
\end{tabularx}
\caption[no caption]{Goals and primary field locations of the five phases of the BSPRSS study.\endnote{The BSPRSS final report combines the first three phases into one phase and instead references Phases I, II, and III.}}
\label{table:1}
\end{table*}

The fiscal year 2015 NDAA requires that gender-neutral occupational standards ``(1) accurately predict performance of actual, regular, and recurring duties of a military occupation; and (2) are applied equitably to measure individual capabilities.''\autocite{levin2014national} But the ACFT has neither been shown to accurately predict performance on actual and recurring duties of a soldier nor is it applied equitably.  In a 2018 RAND study that examined gender-neutral physical standards for ground combat operations, authors explain equitability in context of physical standards stating that ``test validity should not differ among relevant subgroups (such as gender and race), and test scores should be unbiased (i.e., two people who receive the same test score should have the same likelihood of success on the job, regardless of subgroup)''\autocite{RAND2018establishing2} The ACFT fails to meet either of these requirements for a valid, unbiased gender-neutral test.

Statistician George E.~P.~Box 
once famously said ``All models are wrong, some are useful.'' The purpose of this review is to examine the ways in which the Army's Baseline Soldier Physical Readiness Requirements Study (BSPRRS)  model are wrong and whether the model is indeed useful. Primary background material for this analysis includes the BSPRRS final report,\autocite{east2019baseline} the University of Iowa's review of that report,\autocite{malek2020review} and earlier technical reports by the U.S.~Army Public Health Command\autocite{jones2015development} and U.S.~Army Research Institute of Environmental Medicine.\autocite{redmond201development162,redmond201development169,redmond201development1610}

\sparagraph{Warrior tasks and battle drills}%
In an effort to develop an updated and evidence-based fitness assessment, the U.S. Army Research Institute for Environmental Medicine  (USARIEM) conducted a major Army-wide fitness prediction study called the Physical Demands Study.\autocite{jones2015development} As part of their analysis USARIEM identified five domains of combat physical fitness: muscular strength, muscular endurance, aerobic endurance, explosive power, and anaerobic endurance. Using these five domains researchers from the U.S.~Army Center for Initial Military Training (CIMT) conducted Army-wide focus groups to identify eleven physically demanding Warrior Tasks and Battle Drills and Common Soldier Tasks (WTBD/CST), such as move as a member of a team and drag a casualty to immediate safety. These eleven tasks were later distilled down to five criterion task vignettes. 
Soldiers performed a 1.6\,km loaded walk/run (the pre-fatigue)\endnote{This was used to simulate a 10\,km road march intended to replicate the physical pre-fatigue created by moving over uneven terrain to the objective.} and then executed a series of four, timed WTBD simulation test vignettes  (\WTST):

\begin{enumerate}
\item \textit{Build a hasty fighting position.} 
The soldier filled five 5-gallon buckets with sand using an e-tool. After this they carried sixteen 40-lb sandbags, generally  one or two  at a time, ten meters and stacked them onto a platform.
\item \textit{Move over, under, around and through.} 
The soldier completed a course that simulated commonly occurring obstacles in urban/forest terrains: moving 10\,m in a high crawl; zigzag running for 45\,m while jumping over low obstacles, ditches, and tires; traversing a 24' V-shaped balance beam while carrying a squad automatic weapon and an ammo can; lifting a 50-lb rucksack onto the 48'' platform, climbing onto and then moving across the platform, and finally lowering themselves and the object to the ground; scaling a 54'' wall; moving over a 42'' barrier, under an 18'' barrier, through a window, through a 24'' $\times$ 10' tunnel, and over another 42'' barrier; with a combined 50\,m sprinting between events. 
\item \textit{React to man-man contact.} 
The soldier performed a set of four obstacles to simulate the physical demands associated with hand-to-hand contact such as pushing, pulling, grasping, and throwing. These obstacles included flipping a 107-lb tire over four times; pushing a 163-lb prowler sled 20\,m;\endnote{This obstacle was changed from a 20-m power drag of a 163-lb weighted evacuation sled.} lifting and throwing five 30-lb sandbags over a 54'' wall; and rotating a 55-gallon trashcan filled with 300\,lbs of sand two complete turns clockwise and then two complete turns counter-clockwise.
\item \textit{Extract/evacuate a casualty.} 
Starting from a prone position beside a barrier, the soldier stood and rushed to a second barrier where they took a knee, and then they completed a short crouch run to a ``disabled Humvee''\endnote{The researchers originally used an actual Humvee for the casualty-extraction-evacuation vignette. The time to get the combatives dummy back into the Humvee between vignettes was so long as to be impractical that they changed the casualty extraction to a wooden bench seat and later to a flat table.} (a 4' $\times$ 6' plywood platform with a 47'' height and a 2'' border), where the soldier  extricated a 182-lb training dummy and lowered it to the ground. Finally, the soldier dragged the dummy 20\,m to safety and then sprinted 65\,m.
\end{enumerate}

The total time of these four events (not including the pre-fatigue walk/run) became a baseline for physical fitness readiness. One might question whether composite vignette completion time alone is the best indicator of fitness.\endnote{The University of Iowa reviewers did question this metric.} It is not time-on-task that is fundamental to a predictive linear model but the variance of time-on-task among all soldiers. Furthermore, using a composite time instead of individual vignette times will implicitly weight each vignette  by their individual standard deviations.
\endnote{The composite standard deviation will be between the sum of the standard deviations $(L^1)$ and the square root of the sum of the squares of the standard deviations $(L^2)$ depending on the degree of correlation between vignette times. For $L^{1.3}$ the relative importance of each vignette is hasty fighting  (37\%), move OUAT (14\%), combatives (27\%), and casualty evacuation (22\%).}

\sparagraph{Predicting a soldier's physical performance}%
It would be impractical to use the \WTST\ as the fitness test of record for every soldier in the Army, so the researchers looked for possible surrogate physical fitness test events that would be predictive of timed performance on the \WTST\ vignettes. Through a systematic review and the focus group/survey responses, the researchers identified 23 possible predictor test events that would measure muscular strength, explosive power, muscular endurance, cardiovascular endurance, and speed and agility. Many of the tasks of the \WTST\ vignettes required specific physical abilities such as balance, flexibility, and coordination (identified as components of physical fitness by the U.S. Army Center for Health Promotion and Preventive Medicine). For example, traversing the 24-ft V-shaped  beam would require a fair amount of balance. Standard physical tests to measure balance and flexibility like standing balance test and sit-and-reach are noticeably absent from the 23 test event candidates. 
 
The researchers' goal was now to determine which smaller set of these 23 test events would be most predictive of the total time to execute all four vignettes. The hope was that with such a model a soldier would need only their scores from these proxy fitness test events to predict their score on the \WTST\ were they to actually take it. To do this, researchers used a linear regression model. Linear regression is a simple approach and probably an overly simple one: start with a bunch of data points, draw the best straight line\endnote{For multiple regression, the line is a hyperplane.} through those  points, and then use that line to make predictions on future events. In the WTBD prediction model, each of a soldier's test event scores (two-mile run time, number of push-ups, maximum deadlift weight, etc.) are multiplied by a conversion factor to change that score into an equivalent number of seconds. Then all of these seconds are added together, along with some baseline time, for a total time meant to predict the composite time of the \WTST. 
It all sounds rather clever, and the Army makes the extraordinary claim that using this model has an ``80\% ability to predict WTBD/CST performance'' from just the ACFT scores. But, the Army has never validated this claim, the model has a number of serious flaws, and the Army does not implement the ACFT in a manner consistent with the model. 
 
Linear regression models have several requirements. One  is that outcome variables and predictor variables must be linearly correlated. This means that the weighted predictor scores are simply added together to compute a composite score. So, being especially strong or fast on one test event can offset being particularly weak or slow in another one---a fast run time can offset the number of push-ups or leg tucks. And, unlike the ACFT, linear models do not have maximum scores or minimum passing gold standards. Inability to complete one leg tuck is not a failure.  
 
Ideally, the predictor variables in a linear model are largely uncorrelated with themselves. This reduces the number of redundant variables, simplifying and making the model more explainable. Often great care is taken in the design and selection of predictor variables to reduce multicolinearity. Routine exploratory data analysis techniques such as pairwise scatter plots are used to examine data sets for correlation. And statistical techniques, such as principal component analysis, factor analysis, and variance inflation factor comparison, help to quantify  interrelations and identify possible latent variables. Finally, predictive linear regression models  require that the data are representative of the population and are homoscedastic, and that the residuals are normally distributed.\endnote{A least squares regression solution is the maximum likelihood estimator of normally distributed data.} That the Army has been conducting push-ups, sit-ups, and two-mile runs for every soldier over the past forty years should provide researchers ample data with a minimum baseline to compare the test participants.

\sparagraph{Data problems}%
The BSPRRS model had several other compounding data problems. For example, the casualty-extraction-evacuation vignette was performed in as little as 11 seconds and as much as 516 seconds (almost fifty times longer!), with most times around 90 seconds. When a task that typically takes a minute to complete takes over eight minutes, the quality and representativeness of the data should be examined. Because linear regression minimizes the root mean square error, extreme outliers will have a significant impact on the model fit. The researchers assert that \textQT{based on current best practice for regression analyses, individual event scores were not analyzed or adjusted for distribution abnormalities, which is generally considered to be unnecessary with a least squares model (Fox, 2016).} Their statement is simply not true. Chapter~11,  ``Unusual and Influential Data,'' of the cited textbook states emphatically in a bold call-out box centered on the page:  \textQT{Unusual data are problematic in linear models fit by least squares, because they can unduly influence the results of the analysis and because their presence may be a signal that the model fails to capture important characteristics of the data.}\autocite[p.~270]{fox2015applied} When developing any statistical test great care must be taken to understand the data and any outliers that are misrepresentative and would unduly skew the model. 
 
The researchers also state ``for incomplete records with minimal missing data, researchers used mean/linear extrapolation to complete the record.” These extrapolations led to data peculiarities such as a quarter of all female soldiers as being reported as doing a negative number of pull-ups. It is both mathematically and physically impossible to do anything less than zero pull-ups. 
 
Perhaps the most egregious data problem for a study on developing a gender-neutral predictor model is that the data is not representative of all soldiers in the Army. The initial training data consisted of a group of 290 male and 49 female soldiers (mean age of 24 years with a standard deviation of 4.4 years). For comparison, in 2015 the average military officer was roughly 35 years old and the average enlisted member was just over age 27.\autocite{pew2020facts} 
The underrepresentation of women during the development of the model was so significant that the Army researchers stated:  \textQT{following an external review by the University of Iowa, Virtual Soldier Research Center, reviewers suggested we bootstrap additional women into the FT Riley sample to provide a more balanced model and determine if women used a different solution set for \WTST\  performance.} Bootstrapping is a technique where data is resampled from already counted data. In effect the researchers simply copy/pasted already overly underrepresented women, virtually cloning an extra 92 women from the original 49. Unlike in electronic music, resampling  will not create anything fresh. Even worse, the version of the BSPRRS model that the Army touts as having an 80 percent ability to predict WTBD/CST performance was developed using data from a mere 16 women out of 152 total participants.\endnote{Women appear to be overlooked so much in the BSPRR study that while there were several analysis and comparison tables in the final report dedicated only to men, there were no equivalent tables for women.}

\sparagraph{Latent variables}%
The researchers applied the same ideas to develop a predictive model to examine the original APFT (push-ups, sit-ups, and two-mile run). Table~\ref{table:2} shows the linear regression coefficients derived from the training data along with the mean and standard deviations for each of the individual test events for both men and women. 
 
\begin{table}\centering
\begin{tabularx}{\linewidth}{>{\rowmac}l *{4}{>{\rowmac}Y}<{\clearrow}}
\toprule
{event}&{constant}&\makebox[0pt][c]{push-ups}&{sit-ups}&{run}\\
\midrule
coefficient & 329.6 & $-$7.19 & 3.61 & 0.79 \\ 
mean & \NA & 62.8 & 69.3 & 885  \\
SD  & \NA & 15.2 & 10.8 & 91\\\midrule
\end{tabularx}
\caption{Linear regression coefficients from Fort Riley training data along with the mean and standard deviations for each of the individual test events for both men and women.}
\label{table:2}
\end{table}
 
We can interpret the coefficients as conversion factors, changing the number of push-ups, the number of sit-ups, and two-mile run time into their equivalent seconds of \WTST\ composite time. For example, doing one extra push-up is approximately equivalent to subtracting seven seconds from the \WTST, and cutting ten seconds off the two-mile test event is about the same as cutting eight seconds off the \WTST\ time.%
\endnote{Expected value is a linear operator. By taking the dot product of the coefficients and the means and then adding the constant term, we get a total of 831 seconds, which agrees closely with the average composite \WTST\ mean time of 842 seconds. Without knowing more about the correlation of the variables, we can't transform the variances, but we can make an estimate. If the variables were linearly independent, their covariances would be zero and the standard deviation is a weighted $L^2$-norm of standard deviations. If the variables were highly correlated, then the standard deviation would be a weighted $L^1$-norm of the standard deviation of the independent variables. By taking the dot product of the absolute values of the coefficients and the standard deviations, we get a total of 220 seconds. Computing the weighted $L^2$-norm (square root of the dot product of these terms squared) we have 137 seconds. The average composite \WTST\ standard deviation is 234 seconds. The difference between the predicted and the average lies in the explained variation.}
Oddly, in this model, the number of sit-ups counts against a soldier's \WTST\ predictive performance.\endnote{One might question whether there is a typo or a sign error in the table. But, after reconstructing the \WTST\ mean time and standard deviation from those of the predictor variable, it is clear that the numbers in the table are correct.}  Doing one extra sit-up is equivalent to adding 3.6 seconds onto the \WTST\ time. That is to say, the more sit-ups a soldier does, the less fit he or she is perceived as being. Why would the model make such a prediction? It would seem that sit-ups would be a useful measure of core strength and the more sit-ups the more fit, right? Yes, but push-ups, two-mile run, and sit-ups combined are also a predictor of gender. Most men can do more push-ups and can run faster than women. But, men and women are exactly the same on sit-ups. Well, almost. When the researchers tested 278 men and 46 women at Fort Riley, the highest number of sit-ups among men was 102, and the highest among women was 105 (even among a much smaller sample size). According to the Army's model, those 105 sit ups effectively added almost 8 minutes onto this soldier's two-mile run time (or subtracted 52 push-ups). 
 
Statistical models often measure latent or hidden variables. Gender can be a latent variable. When people refer to a machine learning algorithm as sexist or racist, often it's because the algorithm is finding the latent variables of gender or race buried in the data that it scoops up. Likewise, the Army may have inadvertently built a model trained to look for gender. It seems that the researchers may possibly have realized this, but they state that Army senior leaders directed them to continue with a gender-biased model. There is at least one section in the BSPRRS final report where the researchers possibly acknowledged their concern:  \textQT{With direction from Army senior leaders, all regression analyses were conducted on the complete sample (both men and women). The reasoning was that baseline Warrior Tasks and Battle Drills and Common Soldier Tasks are criterion tasks that apply equally to men and women.}

Height and body mass are themselves latent variables of gender. Women in general are shorter and have lower body mass than men. Female soldiers at Fort Riley and Fort Benning were on average five inches shorter and had 37 pounds less body weight. In designing the \WTST\ events, researchers asked participants to evaluate the difficulty of the tasks and stated  \textQT{women's concerns related to effects of height and body mass. Taller/higher body mass Soldiers did not identify the same problems.} The Army operates as squad/buddy teams rather than as individual soldiers. For example, researchers cited virtually all respondents agreed that scaling a two-meter wall in full fighting load was a two-soldier task and that few Soldiers could scale such a wall in full combat load (85\,lbs) as an individual task. In order to design a task that assesses an individual soldier instead of a buddy team, the research team used a modified fighting load weight (50\,lbs) and a 1.4-m wall.\endnote{I am 70 inches tall (the average height of a male soldier participating in the test events). To me a 1.4-\,m wall is the good height to get proper leverage. If I were five inches shorter (the average height of a female soldier), scaling such a wall would be considerably more challenging.} Under this test scenario taller soldiers have an advantage that may not be representative of scaling a two-meter wall as the buddy team. 

When researchers at Fort Carson looked for explicit correlations between weight, height, and fitness variables, they discovered that  height and weight among female soldiers were highly correlated to all \WTST\ vignette times except the move-over-under-around-through-obstacles vignette.\endnote{Reseachers ranked correlation with Pearson correlation values. Height and weight typically scored $R<-0.5$ for female soldiers and $R\approx -0.2$ for male soldiers.} 
Height and weight were both negatively correlated, meaning that taller and heavier female soldiers performed faster on these vignettes. Researchers found no significant correlation among female soldiers for fitness variables (sit-ups, push-ups, and run), and they found no significant correlation between height, weight, or fitness variables and \WTST\ time among male soldiers.\autocite[pp.~13--15]{jones2015development}
When researchers asked one professor of kinesiology whether height and weight affect performance on a proposed fitness assessment, he wrote back in all caps ``\uppercase{this is basic human physiology and so robust as to render it axiomatic.}''
\autocite[p.~E--16]{jones2015development}

Predictive models can reflect biases in subtle ways and when trained on underrepresented groups such models may amplify those biases. 
BSPRRS Phase 4 (Fort Riley) included only 46 women (14.3 percent) and Phase 5 (Fort Benning) included only 16 women (10.5 percent), a lower representation than even the Army. In their review, the University of Iowa Virtual Soldier Research Center criticized the study for the significant under representation of women, finding that due to the  \textQT{inherently unbalanced study design\dots determination of which tasks best predict or represent WTBD/CST performance could be influenced towards strategies used predominantly by men.}\autocite[p.~10]{malek2020review}

\sparagraph{Coefficient of determination}%
When the Army makes claims like ``The APFT is a relatively poor predictor ($\sim$40\%)'' or ``Army Combat Fitness Test (ACFT) $\sim$80\% ability to predict WTBD/CST performance,'' they are referring to the coefficient of determination $R^2$.
The coefficient of determination is a standard measure of the variability of data used to build a model, but it is incorrect to state that $R^2$ is a measure of predictiveness of the model.
Simply stated, $R^2$ is the percentage of variance in the outcome variable that is explainable or accounted for by the predictor variables.\endnote{There are several equivalent definitions for $R^2$ including as the percentage of the explained variance and as the square of the Pearson correlation coefficient between the observed outcome variables and the predicted outcome variables.}
It ranges from zero to one, with $R^2 = 0$ (or zero percent) meaning that there is so much variability in the data that it's impossible to determine any trend, and $R^2 = 1$ (or 100 percent) meaning that all of the data already lines up perfectly and exactly along a line.
So, $R^2 = 0.8$ means that 80 percent of the variance in outcome  can be explained by the variance in the predictor variables. It does not mean predictor variables have an 80 percent ability to predict performance, as the Army has claimed.
\endnote{The BSPRRS final report includes $p$-values when it mentions $R^2$. A $p$-value is the likelihood that the outcome data and the predictor data are uncorrelated, i.e., the likelihood that $R^2$ equals zero or equivalently that all the regression coefficients are zero. In the null hypothesis, the $R^2$ is distributed as a beta distribution $\operatorname{Beta}((k-1)/2,(n-k)/2)$, where $k$ is the number of coefficients including the constant term and $n$ is the sample size.  When $n =  300$ and $k = 6$, we find that even a $p$-value of less than $0.001$ whenever $R^2$ is greater than $0.07$. In other words, the  $p$-value is not a meaningful statistic.}
 Variability in data should be expected, especially when grouping different people together regardless of gender and age. Even among top athletes some are sprinters, some are marathon runners, some are gymnasts—everyone is different. 

\begin{figure}

	\begin{tikzpicture}
	\begin{axis}[width=0.65\linewidth,height=3\bbarsep,scale only axis, xmin=0,xmax=100,ymin=.5,ymax={3+.5}, xtick = {0,50,100},xticklabels= {\small\textcolor{newgray}{0\%},\small\textcolor{newgray}{50\%},	\small\textcolor{newgray}{100\%}},ytick=\empty,clip=false,axis line style={draw=none},tick style={draw=none}]
	\foreach \x in {1,2,...,3}\addplot[newgray,thin,opacity=0.5] coordinates {(0,\x) (100,\x)};
	\input{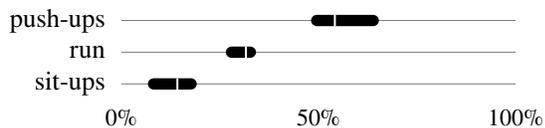}
\end{axis}
\end{tikzpicture}
\caption[no caption]{Estimated relative importance of push-ups, two-mile run, sit-ups in predicting \WTST\ performance as a percentage of the $R^2$. The bars show likely upper and lower bounds of estimation and the notches depict a reasonable point estimate.}
\label{Fig:1}
\end{figure}
While adding more variables into the model will always increase its $R^2$, a more complex model is not necessarily a better one. Therefore, a goal should not necessarily be one that raises the $R^2$ score of a model, a practice often called overfitting. An overfit model may perform wonderfully on the data used to create it, but only on the data used to create it. Once applied to anyone who is not representative of the very narrow training set, the model may be quite wrong. That's precisely what the Army did. The researchers state that the $R^2$ of the original APFT model is 0.423, saying ``the Army Physical Fitness Test (APFT) is a relatively low-to-moderate predictor of WTBD/CSTs performance ($R^2 = 0.423$)\dots demonstrating that the APFT is insufficient to ensure Soldiers are capable of performing physically demanding, commonly occurring, and critical Warrior Tasks and Battle Drills and Common Soldier tasks.” This statement is simply misleading. The linear regression model in and of itself is the predictor of the \WTST\ performance—not the APFT. Also, the \WTST\ composite time is in and of itself a proxy to the WTBD/CST. This is the same model that counts sit-ups against you and the same model that does not account for variability caused by differences in age and gender. 
 
\begin{figure}\centering\setlength{\tabcolsep}{4pt}
\newcommand\anscombe[2]{
\begin{tikzpicture}
\begin{axis}[width=.18\textwidth,height=.18\textwidth,scale only axis,
xmin=0, xmax=20, xtick=\empty, ymin=2, ymax=14, ytick=\empty]
\addplot [fill=black,draw opacity=0, fill opacity =0.75, mark=*,mark size=2pt,only marks,] coordinates{#2};
\addplot [draw,line width=2pt,red] coordinates{(-2,2)(22,14)};
\node[below right,align=left] at  (rel axis cs:0,1) {#1};
\end{axis}\end{tikzpicture}}
\begin{tabular}{cc}
\anscombe{1}{(10.00,8.04)(8.00,6.95)(13.00,7.58)(9.00,8.81)(11.00,8.33)(14.00,9.96)(6.00,7.24)(4.00,4.26)(12.00,10.84)(7.00,4.82)(5.00,5.68)}&
\anscombe{2}{(10.00,9.14)(8.00,8.14)(13.00,8.74)(9.00,8.77)(11.00,9.26)(14.00,8.10)(6.00,6.13)(4.00,3.10)(12.00,9.13)(7.00,7.26)(5.00,4.74)}\\[2pt]
\anscombe{3}{(10.00,7.46)(8.00,6.77)(13.00,12.74)(9.00,7.11)(11.00,7.81)(14.00,8.84)(6.00,6.08)(4.00,5.39)(12.00,8.15)(7.00,6.42)(5.00,5.73)}&
\anscombe{4}{(8.00,6.58)(8.00,5.76)(8.00,7.71)(8.00,8.84)(8.00,8.47)(8.00,7.04)(8.00,5.25)(19.00,12.50)(8.00,5.56)(8.00,7.91)(8.00,6.89)}
\end{tabular}
\caption{Anscombe's quartet. Different sets of eleven data points resulting in the same linear regression model, all with the same $R^2$.}
\label{Fig:2}
\end{figure}
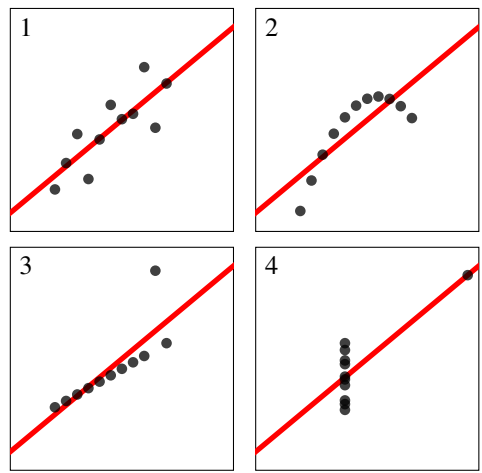

\begin{figure}

	\begin{tikzpicture}
	\begin{axis}[width=0.65\linewidth,height=7\bbarsep,scale only axis, xmin=0,xmax=100,ymin=.5,ymax={7+.5}, xtick = {0,50,100},xticklabels= {\small\textcolor{newgray}{0\%},\small\textcolor{newgray}{50\%},	\small\textcolor{newgray}{100\%}},ytick=\empty,clip=false,axis line style={draw=none},tick style={draw=none}]
	\foreach \x in {1,2,...,7}\addplot[newgray,thin,opacity=0.5] coordinates {(0,\x) (100,\x)};
	\input{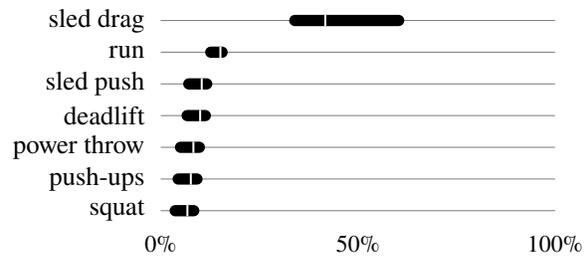}
\end{axis}
\end{tikzpicture}
\caption[no caption]{Estimated relative importance of the seven-test-event Fort Riley model as a percentage of $R^2$.}
\label{Fig:3}
\end{figure}

\begin{figure}

	\begin{tikzpicture}
	\begin{axis}[width=0.65\linewidth,height=8\bbarsep,scale only axis, xmin=0,xmax=100,ymin=.5,ymax={8+.5}, xtick = {0,50,100},xticklabels= {\small\textcolor{newgray}{0\%},\small\textcolor{newgray}{50\%},	\small\textcolor{newgray}{100\%}},ytick=\empty,clip=false,axis line style={draw=none},tick style={draw=none}]
	\foreach \x in {1,2,...,8}\addplot[newgray,thin,opacity=0.5] coordinates {(0,\x) (100,\x)};
	\input{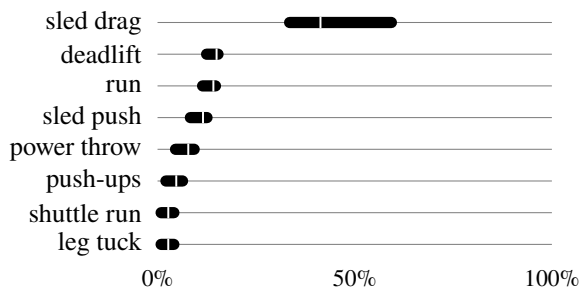}
\end{axis}
\end{tikzpicture}
\caption[no caption]{Estimated relative importance of the adjusted eight-test-event Fort Riley model as a percentage of $R^2$.}
\label{Fig:5}
\end{figure}

\begin{figure}

	\begin{tikzpicture}
	\begin{axis}[width=0.65\linewidth,height=8\bbarsep,scale only axis, xmin=0,xmax=100,ymin=.5,ymax={8+.5}, xtick = {0,50,100},xticklabels= {\small\textcolor{newgray}{0\%},\small\textcolor{newgray}{50\%},	\small\textcolor{newgray}{100\%}},ytick=\empty,clip=false,axis line style={draw=none},tick style={draw=none}]
	\foreach \x in {1,2,...,8}\addplot[newgray,thin,opacity=0.5] coordinates {(0,\x) (100,\x)};
	\input{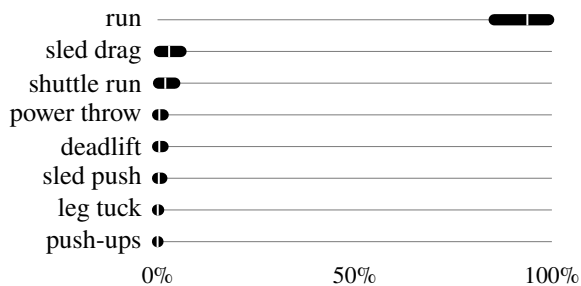}
\end{axis}
\end{tikzpicture}
\caption[no caption]{Estimated relative importance of the eight-test-event Fort Benning model as a percentage of $R^2$.}
\label{Fig:6}
\end{figure}

\begin{figure}

	\begin{tikzpicture}
	\begin{axis}[width=0.65\linewidth,height=6\bbarsep,scale only axis, xmin=0,xmax=100,ymin=.5,ymax={6+.5}, xtick = {0,50,100},xticklabels= {\small\textcolor{newgray}{0\%},\small\textcolor{newgray}{50\%},	\small\textcolor{newgray}{100\%}},ytick=\empty,clip=false,axis line style={draw=none},tick style={draw=none}]
	\foreach \x in {1,2,...,6}\addplot[newgray,thin,opacity=0.5] coordinates {(0,\x) (100,\x)};
	\input{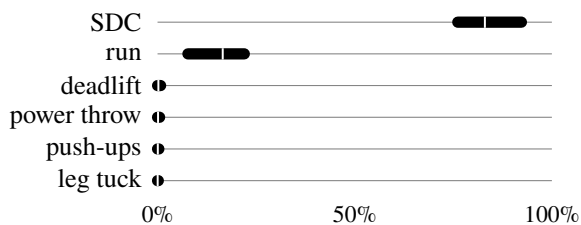}
\end{axis}
\end{tikzpicture}
\caption[no caption]{Estimated relative importance of six-test-event Fort Benning model as a percentage of $R^2$.}
\label{Fig:7}
\end{figure}

Furthermore, a high $R^2$ is not a guarantee that the model is a good representation of the data. Anscombe's quartet is a well-known counter-example that demonstrates that data sampled from very different populations can result in the same predictive linear regression model. See Figure~\ref{Fig:2}. 
The data in each of the four plots is fundamentally different from one another, yet each one has the same mean, variance, and $R^2$.
When these very different data sets are each used to develop predictive linear models, they all result in exactly the same model, all with the same high $R^2 = 0.666$. But, with exception of the first set, none of the fit lines are good models of the data.
The coefficient $R^2$ itself only describes how much of the total variance in the outcome is accounted for by the predictor variables.  It does not tell us how much each predictor variable individually contributes to the outcome variance.  
You should expect that some predictor variables to contribute more than others.
So, a natural question to ask is ``what is the relative importance of each predictor variable in determining $R^2$?''
One would expect that a predictor variable like two-mile-run time or push-up repetitions to have significant relative importance.
Some variables might  have little or no relative importance.
We can always include any variable as a predictor variable.
Preference for peanut butter sandwiches is likely to have zero relative importance as a predictor to the model.
Other variables like gender, age, height, and weight all might have high relative importance, but these variables are not explicitly included in the model.
Parsing the relative importance of each predictor variable also gets complicated, because each variable may likely be correlated with other variables (except  peanut butter sandwich preference,%
but who knows?). 
While it is impossible to compute the relative importance of a test event on \WTST\ outcome without having all the data from each soldier that was used to build the linear predictor model, we can estimate it. Figure~\ref{Fig:1} shows the relative importance of push-ups, run, and sit-ups in determining the performance on the \WTST. Each band shows a likely upper and lower estimate and the notch indicates a reasonable estimate. Technical details are provided as an endnote.%
\endnote{$R^2 = \Var{\sum_{i=1}^n a_i X_i}/{\Var{Y}}$ is the variance in the outcome variables accounted for by the predictor variables divided by the total variance in outcome. 
Let's examine \[S = \Var{\sum_{i=1}^n a_i X_i} = \sum_{i,j=1}^n a_ia_j\Cov{X_i,X_j}.\]
If the variables $X_i$ are all  uncorrelated, then $\Cov{X_i,X_i} = \Var{X_i}$ and $\Cov{X_i,X_j}= 0$ for   $i\neq j$. 
It follows that \[S = \sum_{i=1}^n a_i^2 \Var{X_i} = \sum_{i=1}^n a_i^2 \sigma_i^2,\] and the relative importance is 
\[
\mathrm{RI^2} = \frac{(a_i \sigma_i)^2}{\sum_{i=1}^n (a_i\sigma_i)^2}.
\]
Now, consider the case when the variables $X_i$ were all completely correlated. That is,  Pearson's correlation coefficient $\rho_{ij} = 1$ where the $\rho_{ij} = \Cov{X_i,X_j}/\sigma_i\sigma_j$. It follows that \[S = \Var{\sum_{i=1}^n a_i X_i}  = \left(\sum_{i=1}^n a_i\sigma_i\right)^2.\]
A simple way to partition $S$ is by taking: $S_i =  a_i \sigma_i \sum_{i=1}^n a_i\sigma_i$. In this case, the relative importance is 
\[
\mathrm{RI^1} = \frac{|a_i \sigma_i|}{\sum_{i=1}^n |a_i\sigma_i|}.
\]
If all the terms were completely dependent, linear regression would also be quite ill-conditioned. 
Realistically, variables will have some degree of correlation, but not be completely correlated. Heuristically, a suitable estimate for the relative importance is taking 
\[
\mathrm{RI}^p = \frac{|a_i \sigma_i|^p}{\sum_{i=1}^n |a_i\sigma_i|^p}
\]
for some $1\leq p\leq 2$. Empirically, we can try several values of $p$ to find which one gets $S$ closest to $R^2 \Var{Y}$. We find that $p=1.3$  does a pretty good job, meaning that the variables tend to be a little more linearly dependent than linearly independent. 
The ends of the segments in Figures~\ref{Fig:1} and \ref{Fig:3}--\ref{Fig:7} are determined by computing $\mathrm{RI}^1$ and $\mathrm{RI}^2$ and the notches are determined by computing $\mathrm{RI}^{1.3}$.
(It may happen that terms are anticorrelated, in which case covariance terms are negative, resulting in compounding effects of the variables.  Without the actual data or at very least a covariance matrix, it is difficult or impossible to know such  effects. Covariance matrices are really not difficult to print and should be included!)}

\sparagraph{Fort Riley}
\begin{table*}\centering
\begin{tabularx}{0.9\linewidth}{>{\rowmac}l *{8}{>{\rowmac}Y}<{\clearrow}}
\toprule
event & constant & \makebox[0pt][c]{sled drag} & run & deadlift & \makebox[0pt][c]{sled push} & \makebox[0pt][c]{push-ups} & p.~throw & squat\\\midrule
coefficient & 542.21 & 10.04 & 0.41 & $-$0.60 & 12.29 & $-$1.45 & $-$4.74 & $-$1.45\\
mean  &  *  & 18.3 & 885 & 243.9 & 8.8 & 62.9 & 18.3 & 31.3 \\ 
SD   &   * & 8.1 & 91 & 45.6 & 2.3 & 15.2 & 5.0 & 13.9 \\\midrule
\end{tabularx}
\caption{Regression coefficients for the Fort Riley seven-test-event predictor model.}
\label{table:3}
\end{table*}
\begin{table*}\centering
\begin{tabularx}{1\linewidth}{>{\rowmac}l *{9}{>{\rowmac}Y}<{\clearrow}}
\toprule
event & constant & \makebox[0pt][c]{sled drag} & run & deadlift & \makebox[0pt][c]{sled push} & \makebox[0pt][c]{push-ups} & p.~throw & \makebox[0pt][c]{shuttle run} & leg tuck \\\midrule
coefficient &  436.5 & 9.67 & 0.38 & $-$0.80 & 12.92 & $-$0.98 & $-$4.39 & 1.67 & $-$1.96\\
mean & \NA & 18.3 & 885 & 243.9 & 8.8 & 62.9 & 18.3 & 69.2 & 7.0 \\
SD & \NA & 8.1 & 91 & 45.6 & 2.3 & 15.2 & 5.0 & 5.8 & 5.0\\\midrule
\end{tabularx}
\caption{Regression coefficients and importance for Fort Riley eight-test-event predictor model.}
\label{table:4}
\end{table*}
The CIMT researchers' goal at Fort Riley was to determine which smaller set of the 23 test events (pull-ups, vertical jump, dips, etc.) would be most predictive of the total time to execute all four vignettes and build a predictive linear model using those down-selected test events. Data was collected from a group of 290 male and 49 female soldiers (all volunteers) who performed the \WTST\ vignettes along with all 23 fitness test events over the course of several days.\endnote{The \WTST\ vignettes were conducted three times: a practice trial in combat uniform only with 285 men and 46 women, a trial in modified fighting load without pre-fatigue (a 1.6\,km loaded walk/run with a 55--65-lb load) with 277 men and 44 women, and modified fighting load with pre-fatigue with 256 men and 35 women.  The researchers used the average of the two fighting-load trial times as the dependent variable for the linear regression.  Presumably, participants dropped out of trials and it is unclear how the researchers averaged the missing data or performed regression using missing data. Typically, missing data will need to be removed from the data set. The researchers noted the Pearson correlation factor between the two  modified fighting load  trials was $R = 0.833$. It seems problematic that $R^2 = 0.639$ between the tests is smaller than the $R^2 = 0.737$ of test events. This may indicate overfitting. See page 28 of the BSPRRS final report.}

Through a process called stepwise regression, the researchers successively   either kept or discarded a test event in an attempt to raise $R^2$. The use of stepwise regression has a fair amount of criticism, and many argue that it should only be considered as a first step in model selection. It doesn't really take into account expert opinion, and it may enable statistical analysis misuses such as data dredging. Other statistical tests should be used in combination with stepwise regression including F-testing, principal component analysis, or factor analysis.
 
Through stepwise regression the researchers identified seven test events that raised the $R^2$ from 0.423, corresponding to the three test events of the original APFT, to 0.737. The events included the sled drag, two-mile run, deadlift, sled push, push-up, kettlebell squat, and power throw.%
\endnote{The BSPRRS final report is inconsistent about whether stepwise regression yielded seven or eight test events. Page 28 of the report states: ``The stepwise linear regression model identified eight variables that accounted for a relatively high percentage of explained variance for \WTST\ performance; $R^2 = 0.737$; $p< 0.05$ (see Table 14). The eight variables were: sled drag, power throw, two-mile run, deadlift, sled push, leg tuck, kettlebell squat, and push-up.''  However, Table 14 only includes seven variables and does not include leg tucks. Table 15, which swaps the shuttle run in for the kettlebell squat and includes the leg tuck, states: ``Adjustments made to ensure proper physiologic balance to include anaerobic endurance, core strength training.''  The shuttle run was added to measure anaerobic endurance, and presumably leg tucks were added to measure core strength. Computing the predicted \WTST\ times using the averaged test event scores and coefficients in Tables 14 and 15 of the BSPRRS final report yields 829.5 and 831, respectively. The coefficients are listed with three digits of precision after the decimal, leading to a potential maximum round off error of 1.6. So, the tables are in agreement when Table 14 does not include leg tucks. Furthermore, the level of relative importance of the variables in the seven-test-event predictor model and the eight-event-predictor model are roughly the same. The least important variables in the eight-variable model are the leg tuck and shuttle run. Furthermore, the University of Iowa BSPRSS review report states that the Army initially identified seven test events and later added leg tucks. So, there is strong evidence to conclude that leg tucks were not originally derived through stepwise regression but were later ``forced into the model'' along with the shuttle run.}
The sled drag, sled push, and power throw measure explosive power. The two-mile run measures speed and cardiovascular endurance. The deadlift and squat measure muscular strength. And the push-up measures muscular endurance. Table~\ref{table:3} shows the regression coefficients.\endnote{By taking the dot product of the coefficients and the means added to the constant term, we get a total of 830 seconds, which agrees closely with the average composite \WTST\ time of 842 seconds. 
Taking the dot product of the absolute values of the coefficients and the standard deviations, we get a total of 240 seconds, which agrees closely with the average composite \WTST\ standard deviation of 234 seconds.}

While the researchers provided a list of seven test events, they neglected to state the relative importance of each test event as a predictor of the outcome. Without the original data, it is impossible to precisely determine the relative importance of each event, but it can be estimated using the coefficients of linear regression and the respective standard deviations.\endnote{The relative importance is the percentage contribution of each predictor variable to $R^2$. Even if we don't know the covariance of the variables, we can estimate bounds by using the weighted $L^1$ or $L^2$ norms of the standard deviations.}
 The sled drag is the most significant of the seven predictors of composite WTBD/CST time in the model, and the kettlebell squat is the least. In fact, the sled drag is four times more important than the kettlebell squat as a predictor in the model. The relative importance of the seven test events is shown in Figure~\ref{Fig:3}.
 
 \def\densitysleddrag{\density{sled drag}{12}{(0.063,0.000)(0.063,0.009)(0.067,0.028)(0.070,0.077)(0.073,0.187)(0.076,0.404)(0.079,0.769)(0.083,1.300)(0.086,1.964)(0.089,2.672)(0.092,3.315)(0.096,3.816)(0.099,4.164)(0.102,4.411)(0.105,4.633)(0.109,4.887)(0.112,5.194)(0.115,5.552)(0.118,5.949)(0.122,6.372)(0.125,6.816)(0.128,7.284)(0.131,7.782)(0.135,8.306)(0.138,8.833)(0.141,9.319)(0.144,9.719)(0.148,10.014)(0.151,10.218)(0.154,10.363)(0.157,10.465)(0.161,10.497)(0.164,10.407)(0.167,10.153)(0.170,9.744)(0.174,9.233)(0.177,8.691)(0.180,8.168)(0.183,7.684)(0.187,7.238)(0.190,6.833)(0.193,6.475)(0.196,6.169)(0.200,5.896)(0.203,5.616)(0.206,5.279)(0.209,4.845)(0.212,4.304)(0.216,3.674)(0.219,3.000)(0.222,2.342)(0.225,1.758)(0.229,1.291)(0.232,0.954)(0.235,0.735)(0.238,0.607)(0.242,0.541)(0.245,0.512)(0.248,0.504)(0.251,0.503)(0.255,0.500)(0.258,0.489)(0.261,0.471)(0.264,0.451)(0.268,0.433)(0.271,0.421)(0.274,0.414)(0.277,0.408)(0.281,0.396)(0.284,0.372)(0.287,0.337)(0.290,0.296)(0.294,0.254)(0.297,0.218)(0.300,0.192)(0.303,0.175)(0.307,0.167)(0.310,0.164)(0.313,0.162)(0.316,0.160)(0.320,0.158)(0.323,0.157)(0.326,0.160)(0.329,0.167)(0.333,0.181)(0.336,0.199)(0.339,0.224)(0.342,0.256)(0.345,0.292)(0.349,0.323)(0.352,0.336)(0.355,0.323)(0.358,0.281)(0.362,0.219)(0.365,0.152)(0.368,0.093)(0.371,0.050)(0.375,0.024)(0.378,0.010)(0.381,0.004)(0.384,0.001)(0.384,0.000)}{(0.048,0.000)(0.048,0.005)(0.055,0.012)(0.062,0.026)(0.069,0.051)(0.076,0.093)(0.083,0.160)(0.090,0.259)(0.097,0.392)(0.104,0.558)(0.111,0.746)(0.118,0.942)(0.126,1.128)(0.133,1.286)(0.140,1.408)(0.147,1.490)(0.154,1.542)(0.161,1.574)(0.168,1.601)(0.175,1.634)(0.182,1.680)(0.189,1.740)(0.196,1.814)(0.203,1.899)(0.210,1.992)(0.217,2.090)(0.224,2.192)(0.231,2.298)(0.238,2.408)(0.245,2.524)(0.252,2.644)(0.259,2.767)(0.266,2.887)(0.273,2.999)(0.280,3.098)(0.287,3.179)(0.294,3.241)(0.301,3.289)(0.308,3.325)(0.315,3.354)(0.322,3.377)(0.329,3.392)(0.336,3.392)(0.343,3.371)(0.350,3.323)(0.357,3.247)(0.364,3.148)(0.371,3.033)(0.379,2.909)(0.386,2.787)(0.393,2.669)(0.400,2.557)(0.407,2.453)(0.414,2.354)(0.421,2.260)(0.428,2.173)(0.435,2.092)(0.442,2.019)(0.449,1.953)(0.456,1.889)(0.463,1.825)(0.470,1.754)(0.477,1.674)(0.484,1.582)(0.491,1.478)(0.498,1.364)(0.505,1.243)(0.512,1.118)(0.519,0.996)(0.526,0.879)(0.533,0.774)(0.540,0.682)(0.547,0.606)(0.554,0.545)(0.561,0.497)(0.568,0.457)(0.575,0.423)(0.582,0.391)(0.589,0.360)(0.596,0.329)(0.603,0.300)(0.610,0.274)(0.617,0.253)(0.625,0.237)(0.632,0.225)(0.639,0.217)(0.646,0.208)(0.653,0.196)(0.660,0.180)(0.667,0.159)(0.674,0.134)(0.681,0.107)(0.688,0.080)(0.695,0.057)(0.702,0.038)(0.709,0.024)(0.716,0.014)(0.723,0.008)(0.730,0.004)(0.737,0.002)(0.744,0.001)(0.744,0.000)}}
 \def\densitydeadlift{\density{deadlift}{84}{(0.270,0.000)(0.270,0.001)(0.273,0.003)(0.277,0.007)(0.281,0.016)(0.285,0.032)(0.288,0.061)(0.292,0.106)(0.296,0.172)(0.300,0.261)(0.303,0.368)(0.307,0.487)(0.311,0.605)(0.315,0.710)(0.318,0.790)(0.322,0.842)(0.326,0.865)(0.330,0.869)(0.334,0.867)(0.337,0.871)(0.341,0.898)(0.345,0.959)(0.349,1.063)(0.352,1.215)(0.356,1.413)(0.360,1.650)(0.364,1.914)(0.367,2.191)(0.371,2.472)(0.375,2.752)(0.379,3.033)(0.382,3.323)(0.386,3.627)(0.390,3.946)(0.394,4.270)(0.398,4.583)(0.401,4.868)(0.405,5.115)(0.409,5.322)(0.413,5.503)(0.416,5.676)(0.420,5.858)(0.424,6.056)(0.428,6.260)(0.431,6.451)(0.435,6.606)(0.439,6.707)(0.443,6.753)(0.446,6.757)(0.450,6.736)(0.454,6.708)(0.458,6.683)(0.462,6.658)(0.465,6.625)(0.469,6.570)(0.473,6.482)(0.477,6.353)(0.480,6.184)(0.484,5.976)(0.488,5.735)(0.492,5.469)(0.495,5.191)(0.499,4.914)(0.503,4.652)(0.507,4.411)(0.510,4.191)(0.514,3.982)(0.518,3.773)(0.522,3.555)(0.526,3.324)(0.529,3.081)(0.533,2.828)(0.537,2.569)(0.541,2.307)(0.544,2.045)(0.548,1.789)(0.552,1.545)(0.556,1.323)(0.559,1.129)(0.563,0.970)(0.567,0.848)(0.571,0.761)(0.574,0.704)(0.578,0.668)(0.582,0.642)(0.586,0.612)(0.590,0.572)(0.593,0.515)(0.597,0.443)(0.601,0.362)(0.605,0.280)(0.608,0.204)(0.612,0.139)(0.616,0.089)(0.620,0.054)(0.623,0.030)(0.627,0.016)(0.631,0.008)(0.635,0.004)(0.638,0.001)(0.642,0.001)(0.642,0.000)}{(0.142,0.000)(0.142,0.002)(0.145,0.004)(0.148,0.009)(0.151,0.020)(0.154,0.041)(0.157,0.077)(0.160,0.133)(0.163,0.216)(0.166,0.326)(0.169,0.461)(0.172,0.611)(0.175,0.762)(0.178,0.897)(0.181,1.004)(0.184,1.075)(0.187,1.113)(0.190,1.125)(0.193,1.129)(0.196,1.144)(0.199,1.186)(0.202,1.271)(0.205,1.410)(0.208,1.607)(0.211,1.856)(0.214,2.148)(0.217,2.468)(0.220,2.800)(0.223,3.135)(0.226,3.468)(0.229,3.806)(0.232,4.157)(0.235,4.527)(0.238,4.914)(0.241,5.307)(0.244,5.684)(0.247,6.027)(0.250,6.322)(0.253,6.570)(0.256,6.786)(0.259,6.993)(0.262,7.212)(0.265,7.450)(0.268,7.699)(0.271,7.933)(0.274,8.126)(0.277,8.257)(0.280,8.321)(0.283,8.330)(0.286,8.307)(0.289,8.273)(0.292,8.241)(0.295,8.213)(0.298,8.179)(0.301,8.123)(0.304,8.030)(0.307,7.893)(0.310,7.706)(0.313,7.472)(0.316,7.195)(0.319,6.885)(0.322,6.552)(0.325,6.214)(0.328,5.887)(0.330,5.583)(0.333,5.305)(0.336,5.046)(0.339,4.796)(0.342,4.540)(0.345,4.270)(0.348,3.984)(0.351,3.687)(0.354,3.382)(0.357,3.074)(0.360,2.763)(0.363,2.456)(0.366,2.156)(0.369,1.874)(0.372,1.619)(0.375,1.400)(0.378,1.222)(0.381,1.088)(0.384,0.991)(0.387,0.924)(0.390,0.873)(0.393,0.823)(0.396,0.762)(0.399,0.683)(0.402,0.587)(0.405,0.480)(0.408,0.371)(0.411,0.271)(0.414,0.186)(0.417,0.119)(0.420,0.072)(0.423,0.041)(0.426,0.021)(0.429,0.010)(0.432,0.005)(0.435,0.002)(0.438,0.001)(0.438,0.000)}}
 \def\densityrun{\density{two-mile run}{6}{(0.446,0.000)(0.446,0.002)(0.450,0.008)(0.455,0.020)(0.460,0.049)(0.464,0.104)(0.469,0.198)(0.473,0.341)(0.478,0.532)(0.483,0.762)(0.487,1.016)(0.492,1.280)(0.497,1.548)(0.501,1.821)(0.506,2.100)(0.511,2.379)(0.515,2.644)(0.520,2.884)(0.524,3.091)(0.529,3.275)(0.534,3.457)(0.538,3.663)(0.543,3.917)(0.548,4.223)(0.552,4.569)(0.557,4.926)(0.561,5.272)(0.566,5.602)(0.571,5.934)(0.575,6.285)(0.580,6.648)(0.585,6.981)(0.589,7.224)(0.594,7.341)(0.599,7.336)(0.603,7.252)(0.608,7.131)(0.612,6.990)(0.617,6.813)(0.622,6.576)(0.626,6.280)(0.631,5.949)(0.636,5.614)(0.640,5.292)(0.645,4.979)(0.649,4.662)(0.654,4.333)(0.659,4.002)(0.663,3.684)(0.668,3.393)(0.673,3.128)(0.677,2.878)(0.682,2.619)(0.686,2.332)(0.691,2.006)(0.696,1.649)(0.700,1.286)(0.705,0.949)(0.710,0.670)(0.714,0.467)(0.719,0.342)(0.724,0.286)(0.728,0.281)(0.733,0.307)(0.737,0.343)(0.742,0.373)(0.747,0.387)(0.751,0.382)(0.756,0.364)(0.761,0.341)(0.765,0.320)(0.770,0.305)(0.774,0.293)(0.779,0.280)(0.784,0.261)(0.788,0.233)(0.793,0.200)(0.798,0.166)(0.802,0.138)(0.807,0.118)(0.812,0.108)(0.816,0.107)(0.821,0.114)(0.825,0.129)(0.830,0.150)(0.835,0.173)(0.839,0.190)(0.844,0.196)(0.849,0.190)(0.853,0.174)(0.858,0.151)(0.862,0.124)(0.867,0.097)(0.872,0.071)(0.876,0.049)(0.881,0.030)(0.886,0.017)(0.890,0.008)(0.895,0.004)(0.899,0.001)(0.904,0.000)(0.904,0.000)}{(0.523,0.000)(0.523,0.003)(0.526,0.008)(0.530,0.016)(0.534,0.033)(0.538,0.063)(0.542,0.112)(0.546,0.188)(0.550,0.299)(0.553,0.449)(0.557,0.640)(0.561,0.869)(0.565,1.126)(0.569,1.400)(0.573,1.675)(0.577,1.936)(0.580,2.171)(0.584,2.371)(0.588,2.532)(0.592,2.654)(0.596,2.742)(0.600,2.805)(0.604,2.856)(0.607,2.906)(0.611,2.969)(0.615,3.055)(0.619,3.172)(0.623,3.322)(0.627,3.501)(0.630,3.704)(0.634,3.919)(0.638,4.134)(0.642,4.341)(0.646,4.536)(0.650,4.725)(0.654,4.915)(0.657,5.116)(0.661,5.330)(0.665,5.552)(0.669,5.767)(0.673,5.952)(0.677,6.088)(0.681,6.162)(0.684,6.173)(0.688,6.135)(0.692,6.065)(0.696,5.982)(0.700,5.897)(0.704,5.810)(0.708,5.716)(0.711,5.602)(0.715,5.461)(0.719,5.295)(0.723,5.111)(0.727,4.920)(0.731,4.735)(0.735,4.560)(0.738,4.395)(0.742,4.234)(0.746,4.070)(0.750,3.896)(0.754,3.711)(0.758,3.517)(0.762,3.320)(0.765,3.126)(0.769,2.943)(0.773,2.774)(0.777,2.620)(0.781,2.480)(0.785,2.348)(0.789,2.218)(0.792,2.084)(0.796,1.942)(0.800,1.792)(0.804,1.636)(0.808,1.482)(0.812,1.338)(0.816,1.210)(0.819,1.101)(0.823,1.009)(0.827,0.930)(0.831,0.858)(0.835,0.788)(0.839,0.719)(0.843,0.648)(0.846,0.577)(0.850,0.505)(0.854,0.434)(0.858,0.364)(0.862,0.295)(0.866,0.231)(0.869,0.173)(0.873,0.124)(0.877,0.084)(0.881,0.053)(0.885,0.032)(0.889,0.018)(0.893,0.010)(0.896,0.005)(0.900,0.002)(0.904,0.001)(0.904,0.000)}}
 \def\densitysledpush{\density{sled push}{15}{(0.150,0.000)(0.150,0.029)(0.156,0.108)(0.161,0.326)(0.167,0.797)(0.173,1.577)(0.179,2.553)(0.184,3.421)(0.190,3.890)(0.196,3.941)(0.202,3.832)(0.207,3.860)(0.213,4.124)(0.219,4.535)(0.225,4.958)(0.230,5.334)(0.236,5.678)(0.242,6.025)(0.248,6.377)(0.253,6.712)(0.259,7.010)(0.265,7.260)(0.271,7.431)(0.276,7.470)(0.282,7.353)(0.288,7.115)(0.294,6.806)(0.299,6.459)(0.305,6.095)(0.311,5.745)(0.317,5.422)(0.322,5.082)(0.328,4.640)(0.334,4.038)(0.340,3.301)(0.345,2.519)(0.351,1.802)(0.357,1.231)(0.362,0.834)(0.368,0.580)(0.374,0.425)(0.380,0.327)(0.385,0.262)(0.391,0.218)(0.397,0.190)(0.403,0.174)(0.408,0.166)(0.414,0.159)(0.420,0.146)(0.426,0.128)(0.431,0.109)(0.437,0.100)(0.443,0.107)(0.449,0.129)(0.454,0.159)(0.460,0.183)(0.466,0.193)(0.472,0.187)(0.477,0.172)(0.483,0.157)(0.489,0.146)(0.495,0.144)(0.500,0.148)(0.506,0.148)(0.512,0.134)(0.518,0.108)(0.523,0.085)(0.529,0.080)(0.535,0.096)(0.541,0.123)(0.546,0.142)(0.552,0.141)(0.558,0.120)(0.564,0.090)(0.569,0.063)(0.575,0.047)(0.581,0.044)(0.587,0.049)(0.592,0.055)(0.598,0.058)(0.604,0.058)(0.610,0.060)(0.615,0.070)(0.621,0.095)(0.627,0.134)(0.633,0.182)(0.638,0.227)(0.644,0.254)(0.650,0.251)(0.656,0.221)(0.661,0.177)(0.667,0.137)(0.673,0.108)(0.679,0.086)(0.684,0.066)(0.690,0.045)(0.696,0.026)(0.702,0.013)(0.707,0.005)(0.713,0.002)(0.719,0.000)(0.719,0.000)}{(0.243,0.000)(0.243,0.018)(0.246,0.036)(0.250,0.068)(0.254,0.124)(0.258,0.213)(0.262,0.346)(0.266,0.534)(0.270,0.781)(0.274,1.085)(0.278,1.432)(0.282,1.796)(0.285,2.144)(0.289,2.443)(0.293,2.665)(0.297,2.793)(0.301,2.830)(0.305,2.795)(0.309,2.719)(0.313,2.635)(0.317,2.574)(0.321,2.555)(0.324,2.587)(0.328,2.666)(0.332,2.782)(0.336,2.919)(0.340,3.064)(0.344,3.205)(0.348,3.336)(0.352,3.455)(0.356,3.564)(0.360,3.668)(0.363,3.771)(0.367,3.876)(0.371,3.987)(0.375,4.101)(0.379,4.218)(0.383,4.333)(0.387,4.444)(0.391,4.548)(0.395,4.644)(0.399,4.733)(0.402,4.817)(0.406,4.894)(0.410,4.965)(0.414,5.023)(0.418,5.064)(0.422,5.082)(0.426,5.072)(0.430,5.035)(0.434,4.975)(0.437,4.896)(0.441,4.807)(0.445,4.712)(0.449,4.613)(0.453,4.510)(0.457,4.403)(0.461,4.290)(0.465,4.172)(0.469,4.054)(0.473,3.939)(0.476,3.832)(0.480,3.735)(0.484,3.644)(0.488,3.555)(0.492,3.460)(0.496,3.349)(0.500,3.218)(0.504,3.061)(0.508,2.881)(0.512,2.682)(0.515,2.472)(0.519,2.261)(0.523,2.060)(0.527,1.878)(0.531,1.721)(0.535,1.593)(0.539,1.493)(0.543,1.419)(0.547,1.364)(0.551,1.321)(0.554,1.281)(0.558,1.235)(0.562,1.178)(0.566,1.103)(0.570,1.011)(0.574,0.901)(0.578,0.780)(0.582,0.652)(0.586,0.525)(0.590,0.407)(0.593,0.302)(0.597,0.215)(0.601,0.146)(0.605,0.095)(0.609,0.059)(0.613,0.035)(0.617,0.019)(0.621,0.010)(0.625,0.005)(0.628,0.003)(0.628,0.000)}}
 \def\densitypushups{\density{push-ups}{10}{(0.135,0.000)(0.135,0.002)(0.140,0.007)(0.144,0.019)(0.149,0.045)(0.153,0.095)(0.157,0.181)(0.162,0.314)(0.166,0.493)(0.171,0.710)(0.175,0.944)(0.180,1.170)(0.184,1.372)(0.189,1.543)(0.193,1.690)(0.198,1.828)(0.202,1.971)(0.207,2.128)(0.211,2.306)(0.216,2.506)(0.220,2.732)(0.225,2.989)(0.229,3.285)(0.234,3.626)(0.238,4.009)(0.243,4.421)(0.247,4.843)(0.252,5.251)(0.256,5.631)(0.261,5.979)(0.265,6.294)(0.270,6.572)(0.274,6.797)(0.279,6.944)(0.283,6.996)(0.288,6.954)(0.292,6.841)(0.297,6.698)(0.301,6.561)(0.306,6.444)(0.310,6.341)(0.315,6.230)(0.319,6.089)(0.324,5.915)(0.328,5.721)(0.332,5.526)(0.337,5.345)(0.341,5.173)(0.346,4.990)(0.350,4.767)(0.355,4.483)(0.359,4.127)(0.364,3.711)(0.368,3.257)(0.373,2.797)(0.377,2.360)(0.382,1.963)(0.386,1.616)(0.391,1.320)(0.395,1.075)(0.400,0.879)(0.404,0.731)(0.409,0.629)(0.413,0.567)(0.418,0.535)(0.422,0.521)(0.427,0.512)(0.431,0.500)(0.436,0.479)(0.440,0.450)(0.445,0.415)(0.449,0.377)(0.454,0.340)(0.458,0.307)(0.463,0.278)(0.467,0.252)(0.472,0.228)(0.476,0.205)(0.481,0.184)(0.485,0.169)(0.490,0.161)(0.494,0.159)(0.499,0.165)(0.503,0.176)(0.507,0.190)(0.512,0.205)(0.516,0.216)(0.521,0.220)(0.525,0.215)(0.530,0.199)(0.534,0.174)(0.539,0.142)(0.543,0.108)(0.548,0.076)(0.552,0.049)(0.557,0.029)(0.561,0.015)(0.566,0.008)(0.570,0.003)(0.575,0.001)(0.579,0.000)(0.579,0.000)}{(0.048,0.000)(0.048,0.022)(0.053,0.055)(0.057,0.125)(0.061,0.258)(0.066,0.480)(0.070,0.812)(0.074,1.248)(0.079,1.750)(0.083,2.247)(0.087,2.661)(0.092,2.936)(0.096,3.063)(0.100,3.082)(0.105,3.064)(0.109,3.081)(0.113,3.178)(0.118,3.373)(0.122,3.654)(0.127,3.996)(0.131,4.370)(0.135,4.748)(0.140,5.109)(0.144,5.443)(0.148,5.749)(0.153,6.027)(0.157,6.273)(0.161,6.477)(0.166,6.621)(0.170,6.688)(0.174,6.675)(0.179,6.595)(0.183,6.476)(0.187,6.348)(0.192,6.233)(0.196,6.136)(0.200,6.045)(0.205,5.943)(0.209,5.816)(0.213,5.661)(0.218,5.489)(0.222,5.316)(0.226,5.153)(0.231,4.999)(0.235,4.843)(0.239,4.664)(0.244,4.441)(0.248,4.162)(0.253,3.825)(0.257,3.444)(0.261,3.039)(0.266,2.634)(0.270,2.248)(0.274,1.897)(0.279,1.586)(0.283,1.318)(0.287,1.091)(0.292,0.905)(0.296,0.758)(0.300,0.651)(0.305,0.579)(0.309,0.536)(0.313,0.514)(0.318,0.503)(0.322,0.494)(0.326,0.481)(0.331,0.461)(0.335,0.434)(0.339,0.402)(0.344,0.368)(0.348,0.335)(0.352,0.304)(0.357,0.277)(0.361,0.252)(0.365,0.229)(0.370,0.208)(0.374,0.188)(0.378,0.172)(0.383,0.161)(0.387,0.156)(0.392,0.157)(0.396,0.163)(0.400,0.174)(0.405,0.187)(0.409,0.200)(0.413,0.209)(0.418,0.213)(0.422,0.209)(0.426,0.195)(0.431,0.174)(0.435,0.147)(0.439,0.116)(0.444,0.086)(0.448,0.059)(0.452,0.038)(0.457,0.022)(0.461,0.012)(0.465,0.006)(0.470,0.003)(0.474,0.001)(0.478,0.000)(0.478,0.000)}}
 \def\densitypthrow{\density{power throw}{37}{(0.101,0.000)(0.101,0.001)(0.105,0.002)(0.110,0.006)(0.114,0.015)(0.119,0.034)(0.124,0.067)(0.128,0.121)(0.133,0.200)(0.137,0.300)(0.142,0.413)(0.147,0.523)(0.151,0.619)(0.156,0.693)(0.160,0.750)(0.165,0.801)(0.170,0.862)(0.174,0.942)(0.179,1.046)(0.183,1.177)(0.188,1.340)(0.193,1.541)(0.197,1.784)(0.202,2.067)(0.206,2.380)(0.211,2.706)(0.215,3.029)(0.220,3.333)(0.225,3.611)(0.229,3.857)(0.234,4.067)(0.238,4.239)(0.243,4.375)(0.248,4.495)(0.252,4.630)(0.257,4.814)(0.261,5.077)(0.266,5.424)(0.271,5.835)(0.275,6.267)(0.280,6.672)(0.284,7.007)(0.289,7.243)(0.294,7.367)(0.298,7.375)(0.303,7.269)(0.307,7.056)(0.312,6.750)(0.317,6.378)(0.321,5.980)(0.326,5.602)(0.330,5.273)(0.335,5.002)(0.340,4.774)(0.344,4.558)(0.349,4.328)(0.353,4.071)(0.358,3.782)(0.362,3.466)(0.367,3.124)(0.372,2.756)(0.376,2.368)(0.381,1.973)(0.385,1.593)(0.390,1.252)(0.395,0.971)(0.399,0.758)(0.404,0.610)(0.408,0.516)(0.413,0.465)(0.418,0.441)(0.422,0.432)(0.427,0.426)(0.431,0.415)(0.436,0.396)(0.441,0.374)(0.445,0.353)(0.450,0.340)(0.454,0.338)(0.459,0.348)(0.464,0.364)(0.468,0.383)(0.473,0.399)(0.477,0.406)(0.482,0.400)(0.487,0.381)(0.491,0.349)(0.496,0.309)(0.500,0.268)(0.505,0.230)(0.510,0.196)(0.514,0.164)(0.519,0.133)(0.523,0.101)(0.528,0.070)(0.532,0.045)(0.537,0.026)(0.542,0.013)(0.546,0.006)(0.551,0.003)(0.555,0.001)(0.555,0.000)}{(0.060,0.000)(0.060,0.029)(0.062,0.060)(0.065,0.116)(0.067,0.211)(0.069,0.364)(0.071,0.590)(0.073,0.904)(0.075,1.307)(0.077,1.788)(0.080,2.317)(0.082,2.850)(0.084,3.338)(0.086,3.740)(0.088,4.030)(0.090,4.211)(0.093,4.303)(0.095,4.344)(0.097,4.377)(0.099,4.433)(0.101,4.535)(0.103,4.686)(0.105,4.879)(0.108,5.102)(0.110,5.338)(0.112,5.570)(0.114,5.785)(0.116,5.974)(0.118,6.131)(0.120,6.259)(0.123,6.364)(0.125,6.462)(0.127,6.571)(0.129,6.712)(0.131,6.903)(0.133,7.157)(0.135,7.479)(0.138,7.860)(0.140,8.284)(0.142,8.729)(0.144,9.168)(0.146,9.577)(0.148,9.937)(0.151,10.232)(0.153,10.455)(0.155,10.603)(0.157,10.674)(0.159,10.670)(0.161,10.591)(0.163,10.439)(0.166,10.215)(0.168,9.925)(0.170,9.577)(0.172,9.188)(0.174,8.780)(0.176,8.376)(0.178,7.998)(0.181,7.663)(0.183,7.376)(0.185,7.134)(0.187,6.923)(0.189,6.725)(0.191,6.523)(0.193,6.302)(0.196,6.057)(0.198,5.790)(0.200,5.506)(0.202,5.214)(0.204,4.921)(0.206,4.634)(0.209,4.352)(0.211,4.076)(0.213,3.803)(0.215,3.532)(0.217,3.268)(0.219,3.014)(0.221,2.778)(0.224,2.567)(0.226,2.385)(0.228,2.231)(0.230,2.102)(0.232,1.990)(0.234,1.882)(0.236,1.769)(0.239,1.640)(0.241,1.491)(0.243,1.321)(0.245,1.134)(0.247,0.941)(0.249,0.750)(0.251,0.574)(0.254,0.420)(0.256,0.294)(0.258,0.196)(0.260,0.124)(0.262,0.075)(0.264,0.043)(0.266,0.023)(0.269,0.012)(0.271,0.006)(0.273,0.003)(0.273,0.000)}}
 \def\densitysquat{\density{squat}{5}{(-0.019,0.000)(-0.019,0.002)(-0.014,0.006)(-0.009,0.017)(-0.004,0.044)(0.001,0.100)(0.006,0.199)(0.011,0.348)(0.016,0.542)(0.021,0.759)(0.026,0.977)(0.031,1.179)(0.035,1.364)(0.040,1.544)(0.045,1.734)(0.050,1.943)(0.055,2.175)(0.060,2.427)(0.065,2.698)(0.070,2.982)(0.075,3.276)(0.080,3.576)(0.085,3.881)(0.090,4.195)(0.095,4.528)(0.100,4.888)(0.105,5.281)(0.109,5.691)(0.114,6.088)(0.119,6.436)(0.124,6.710)(0.129,6.907)(0.134,7.029)(0.139,7.082)(0.144,7.074)(0.149,7.017)(0.154,6.933)(0.159,6.839)(0.164,6.737)(0.169,6.614)(0.174,6.445)(0.179,6.220)(0.183,5.940)(0.188,5.613)(0.193,5.242)(0.198,4.820)(0.203,4.343)(0.208,3.823)(0.213,3.292)(0.218,2.790)(0.223,2.347)(0.228,1.966)(0.233,1.632)(0.238,1.330)(0.243,1.053)(0.248,0.810)(0.253,0.612)(0.258,0.467)(0.262,0.369)(0.267,0.306)(0.272,0.263)(0.277,0.229)(0.282,0.198)(0.287,0.171)(0.292,0.148)(0.297,0.132)(0.302,0.123)(0.307,0.120)(0.312,0.120)(0.317,0.122)(0.322,0.127)(0.327,0.137)(0.332,0.150)(0.336,0.164)(0.341,0.177)(0.346,0.190)(0.351,0.203)(0.356,0.218)(0.361,0.236)(0.366,0.254)(0.371,0.265)(0.376,0.262)(0.381,0.242)(0.386,0.210)(0.391,0.174)(0.396,0.143)(0.401,0.120)(0.406,0.105)(0.410,0.096)(0.415,0.091)(0.420,0.088)(0.425,0.083)(0.430,0.076)(0.435,0.063)(0.440,0.047)(0.445,0.031)(0.450,0.018)(0.455,0.009)(0.460,0.004)(0.465,0.001)(0.470,0.000)(0.470,0.000)}{(-0.011,0.000)(-0.011,0.021)(-0.009,0.052)(-0.006,0.119)(-0.003,0.247)(-0.001,0.471)(0.002,0.824)(0.004,1.324)(0.007,1.957)(0.009,2.669)(0.012,3.374)(0.014,3.977)(0.017,4.410)(0.020,4.657)(0.022,4.756)(0.025,4.782)(0.027,4.811)(0.030,4.897)(0.032,5.063)(0.035,5.304)(0.037,5.600)(0.040,5.933)(0.043,6.292)(0.045,6.673)(0.048,7.079)(0.050,7.510)(0.053,7.963)(0.055,8.422)(0.058,8.866)(0.060,9.270)(0.063,9.617)(0.065,9.899)(0.068,10.114)(0.071,10.265)(0.073,10.357)(0.076,10.392)(0.078,10.377)(0.081,10.320)(0.083,10.236)(0.086,10.136)(0.088,10.030)(0.091,9.919)(0.094,9.794)(0.096,9.641)(0.099,9.447)(0.101,9.203)(0.104,8.912)(0.106,8.577)(0.109,8.204)(0.111,7.794)(0.114,7.340)(0.117,6.839)(0.119,6.290)(0.122,5.705)(0.124,5.107)(0.127,4.525)(0.129,3.985)(0.132,3.505)(0.134,3.086)(0.137,2.720)(0.139,2.393)(0.142,2.091)(0.145,1.807)(0.147,1.542)(0.150,1.300)(0.152,1.088)(0.155,0.907)(0.157,0.759)(0.160,0.641)(0.162,0.548)(0.165,0.479)(0.168,0.431)(0.170,0.402)(0.173,0.392)(0.175,0.399)(0.178,0.419)(0.180,0.450)(0.183,0.487)(0.185,0.526)(0.188,0.559)(0.191,0.581)(0.193,0.587)(0.196,0.572)(0.198,0.537)(0.201,0.488)(0.203,0.430)(0.206,0.371)(0.208,0.316)(0.211,0.266)(0.213,0.221)(0.216,0.179)(0.219,0.141)(0.221,0.106)(0.224,0.075)(0.226,0.049)(0.229,0.030)(0.231,0.017)(0.234,0.009)(0.236,0.004)(0.239,0.002)(0.242,0.001)(0.242,0.000)}}
 \def\densitylegtuck{\density{leg tuck}{10}{(-0.036,0.000)(-0.036,0.018)(-0.032,0.044)(-0.028,0.102)(-0.024,0.214)(-0.021,0.406)(-0.017,0.700)(-0.013,1.099)(-0.009,1.572)(-0.006,2.057)(-0.002,2.474)(0.002,2.753)(0.006,2.866)(0.010,2.838)(0.013,2.735)(0.017,2.640)(0.021,2.616)(0.025,2.699)(0.029,2.889)(0.032,3.165)(0.036,3.493)(0.040,3.839)(0.044,4.172)(0.048,4.470)(0.051,4.727)(0.055,4.948)(0.059,5.152)(0.063,5.360)(0.067,5.594)(0.070,5.865)(0.074,6.173)(0.078,6.505)(0.082,6.838)(0.086,7.147)(0.089,7.405)(0.093,7.597)(0.097,7.710)(0.101,7.745)(0.105,7.703)(0.108,7.590)(0.112,7.413)(0.116,7.176)(0.120,6.888)(0.124,6.554)(0.127,6.186)(0.131,5.796)(0.135,5.406)(0.139,5.039)(0.143,4.713)(0.146,4.435)(0.150,4.200)(0.154,3.990)(0.158,3.792)(0.162,3.593)(0.165,3.392)(0.169,3.188)(0.173,2.980)(0.177,2.762)(0.181,2.527)(0.184,2.276)(0.188,2.012)(0.192,1.745)(0.196,1.489)(0.200,1.252)(0.203,1.044)(0.207,0.868)(0.211,0.726)(0.215,0.612)(0.219,0.519)(0.222,0.441)(0.226,0.370)(0.230,0.306)(0.234,0.251)(0.238,0.208)(0.241,0.180)(0.245,0.169)(0.249,0.175)(0.253,0.192)(0.257,0.216)(0.260,0.241)(0.264,0.264)(0.268,0.285)(0.272,0.305)(0.275,0.324)(0.279,0.344)(0.283,0.361)(0.287,0.371)(0.291,0.370)(0.294,0.357)(0.298,0.329)(0.302,0.288)(0.306,0.237)(0.310,0.183)(0.313,0.131)(0.317,0.087)(0.321,0.053)(0.325,0.030)(0.329,0.015)(0.332,0.007)(0.336,0.003)(0.340,0.001)(0.340,0.000)}{(-0.013,0.000)(-0.013,0.245)(-0.012,0.543)(-0.010,1.120)(-0.009,2.149)(-0.008,3.839)(-0.007,6.384)(-0.006,9.888)(-0.005,14.273)(-0.003,19.222)(-0.002,24.179)(-0.001,28.461)(-0.000,31.430)(0.001,32.684)(0.002,32.180)(0.003,30.234)(0.005,27.407)(0.006,24.323)(0.007,21.506)(0.008,19.283)(0.009,17.769)(0.010,16.922)(0.012,16.609)(0.013,16.675)(0.014,16.976)(0.015,17.391)(0.016,17.825)(0.017,18.204)(0.018,18.472)(0.020,18.592)(0.021,18.546)(0.022,18.335)(0.023,17.973)(0.024,17.480)(0.025,16.875)(0.026,16.177)(0.028,15.408)(0.029,14.593)(0.030,13.770)(0.031,12.979)(0.032,12.259)(0.033,11.636)(0.035,11.115)(0.036,10.679)(0.037,10.296)(0.038,9.930)(0.039,9.555)(0.040,9.164)(0.041,8.761)(0.043,8.359)(0.044,7.966)(0.045,7.579)(0.046,7.186)(0.047,6.771)(0.048,6.321)(0.050,5.836)(0.051,5.326)(0.052,4.808)(0.053,4.301)(0.054,3.817)(0.055,3.368)(0.056,2.958)(0.058,2.591)(0.059,2.269)(0.060,1.988)(0.061,1.743)(0.062,1.527)(0.063,1.331)(0.065,1.151)(0.066,0.986)(0.067,0.843)(0.068,0.728)(0.069,0.649)(0.070,0.610)(0.071,0.609)(0.073,0.641)(0.074,0.695)(0.075,0.761)(0.076,0.830)(0.077,0.899)(0.078,0.965)(0.080,1.026)(0.081,1.081)(0.082,1.124)(0.083,1.149)(0.084,1.148)(0.085,1.115)(0.086,1.046)(0.088,0.944)(0.089,0.815)(0.090,0.671)(0.091,0.524)(0.092,0.388)(0.093,0.271)(0.094,0.178)(0.096,0.110)(0.097,0.063)(0.098,0.034)(0.099,0.017)(0.100,0.008)(0.101,0.004)(0.101,0.000)}}
 \def\densitysitups{\density{sit-ups}{1}{(0.170,0.000)(0.170,0.001)(0.174,0.002)(0.178,0.005)(0.182,0.011)(0.186,0.023)(0.190,0.042)(0.194,0.070)(0.198,0.108)(0.202,0.156)(0.206,0.211)(0.210,0.270)(0.214,0.328)(0.218,0.383)(0.221,0.430)(0.225,0.467)(0.229,0.494)(0.233,0.513)(0.237,0.531)(0.241,0.557)(0.245,0.601)(0.249,0.672)(0.253,0.776)(0.257,0.917)(0.261,1.100)(0.265,1.325)(0.269,1.591)(0.272,1.896)(0.276,2.235)(0.280,2.597)(0.284,2.976)(0.288,3.362)(0.292,3.746)(0.296,4.123)(0.300,4.487)(0.304,4.834)(0.308,5.159)(0.312,5.462)(0.316,5.740)(0.320,5.986)(0.323,6.192)(0.327,6.352)(0.331,6.466)(0.335,6.540)(0.339,6.591)(0.343,6.640)(0.347,6.708)(0.351,6.802)(0.355,6.914)(0.359,7.021)(0.363,7.098)(0.367,7.124)(0.371,7.096)(0.374,7.022)(0.378,6.914)(0.382,6.776)(0.386,6.600)(0.390,6.373)(0.394,6.086)(0.398,5.741)(0.402,5.359)(0.406,4.970)(0.410,4.601)(0.414,4.270)(0.418,3.978)(0.422,3.712)(0.426,3.449)(0.429,3.171)(0.433,2.864)(0.437,2.531)(0.441,2.188)(0.445,1.857)(0.449,1.560)(0.453,1.313)(0.457,1.123)(0.461,0.985)(0.465,0.893)(0.469,0.832)(0.473,0.793)(0.477,0.764)(0.480,0.738)(0.484,0.709)(0.488,0.674)(0.492,0.633)(0.496,0.587)(0.500,0.538)(0.504,0.489)(0.508,0.442)(0.512,0.395)(0.516,0.347)(0.520,0.296)(0.524,0.241)(0.528,0.185)(0.531,0.132)(0.535,0.088)(0.539,0.053)(0.543,0.029)(0.547,0.015)(0.551,0.007)(0.555,0.003)(0.559,0.001)(0.559,0.000)}{(0.188,0.000)(0.188,0.006)(0.192,0.014)(0.196,0.031)(0.200,0.062)(0.204,0.117)(0.208,0.206)(0.212,0.337)(0.216,0.516)(0.220,0.740)(0.224,0.996)(0.228,1.263)(0.232,1.517)(0.236,1.738)(0.240,1.917)(0.244,2.056)(0.248,2.173)(0.252,2.287)(0.256,2.419)(0.260,2.583)(0.264,2.782)(0.268,3.013)(0.272,3.265)(0.275,3.528)(0.279,3.791)(0.283,4.045)(0.287,4.286)(0.291,4.512)(0.295,4.720)(0.299,4.909)(0.303,5.076)(0.307,5.217)(0.311,5.329)(0.315,5.412)(0.319,5.468)(0.323,5.506)(0.327,5.537)(0.331,5.573)(0.335,5.624)(0.339,5.692)(0.343,5.773)(0.347,5.857)(0.351,5.927)(0.355,5.972)(0.359,5.983)(0.363,5.960)(0.367,5.909)(0.371,5.836)(0.375,5.747)(0.378,5.640)(0.382,5.507)(0.386,5.341)(0.390,5.137)(0.394,4.895)(0.398,4.626)(0.402,4.345)(0.406,4.068)(0.410,3.809)(0.414,3.575)(0.418,3.366)(0.422,3.175)(0.426,2.992)(0.430,2.804)(0.434,2.603)(0.438,2.384)(0.442,2.149)(0.446,1.907)(0.450,1.669)(0.454,1.448)(0.458,1.254)(0.462,1.093)(0.466,0.966)(0.470,0.870)(0.474,0.802)(0.478,0.754)(0.481,0.721)(0.485,0.695)(0.489,0.673)(0.493,0.651)(0.497,0.625)(0.501,0.595)(0.505,0.561)(0.509,0.524)(0.513,0.485)(0.517,0.445)(0.521,0.406)(0.525,0.366)(0.529,0.327)(0.533,0.285)(0.537,0.241)(0.541,0.197)(0.545,0.153)(0.549,0.113)(0.553,0.079)(0.557,0.052)(0.561,0.032)(0.565,0.018)(0.569,0.010)(0.573,0.005)(0.577,0.002)(0.581,0.001)(0.581,0.000)}}
 \def\densityhastyfighting{\density{hasty fighting}{12}{(0.233,0.000)(0.233,0.001)(0.237,0.006)(0.242,0.021)(0.247,0.057)(0.252,0.124)(0.256,0.220)(0.261,0.324)(0.266,0.415)(0.271,0.481)(0.275,0.534)(0.280,0.589)(0.285,0.664)(0.290,0.779)(0.294,0.942)(0.299,1.139)(0.304,1.349)(0.309,1.574)(0.313,1.848)(0.318,2.205)(0.323,2.656)(0.328,3.184)(0.332,3.773)(0.337,4.419)(0.342,5.107)(0.347,5.774)(0.351,6.334)(0.356,6.763)(0.361,7.161)(0.366,7.664)(0.370,8.275)(0.375,8.820)(0.380,9.091)(0.385,9.065)(0.390,8.901)(0.394,8.796)(0.399,8.800)(0.404,8.793)(0.409,8.604)(0.413,8.135)(0.418,7.438)(0.423,6.648)(0.428,5.891)(0.432,5.219)(0.437,4.615)(0.442,4.047)(0.447,3.479)(0.451,2.890)(0.456,2.284)(0.461,1.701)(0.466,1.194)(0.470,0.809)(0.475,0.567)(0.480,0.451)(0.485,0.412)(0.489,0.395)(0.494,0.369)(0.499,0.329)(0.504,0.289)(0.508,0.263)(0.513,0.251)(0.518,0.241)(0.523,0.219)(0.527,0.184)(0.532,0.146)(0.537,0.116)(0.542,0.102)(0.546,0.112)(0.551,0.147)(0.556,0.200)(0.561,0.252)(0.565,0.286)(0.570,0.292)(0.575,0.275)(0.580,0.253)(0.584,0.237)(0.589,0.228)(0.594,0.227)(0.599,0.238)(0.604,0.259)(0.608,0.283)(0.613,0.301)(0.618,0.309)(0.623,0.304)(0.627,0.286)(0.632,0.259)(0.637,0.233)(0.642,0.212)(0.646,0.194)(0.651,0.177)(0.656,0.165)(0.661,0.161)(0.665,0.162)(0.670,0.156)(0.675,0.134)(0.680,0.097)(0.684,0.058)(0.689,0.028)(0.694,0.011)(0.699,0.003)(0.703,0.001)(0.703,0.000)}{(0.346,0.000)(0.346,0.006)(0.352,0.014)(0.358,0.028)(0.363,0.055)(0.369,0.100)(0.374,0.170)(0.380,0.272)(0.386,0.411)(0.391,0.583)(0.397,0.781)(0.402,0.988)(0.408,1.185)(0.414,1.353)(0.419,1.478)(0.425,1.557)(0.430,1.596)(0.436,1.611)(0.442,1.620)(0.447,1.641)(0.453,1.688)(0.458,1.766)(0.464,1.875)(0.470,2.011)(0.475,2.166)(0.481,2.330)(0.486,2.493)(0.492,2.645)(0.497,2.780)(0.503,2.897)(0.509,2.997)(0.514,3.089)(0.520,3.182)(0.525,3.288)(0.531,3.411)(0.537,3.549)(0.542,3.693)(0.548,3.828)(0.553,3.940)(0.559,4.015)(0.565,4.048)(0.570,4.041)(0.576,4.006)(0.581,3.955)(0.587,3.906)(0.593,3.869)(0.598,3.849)(0.604,3.845)(0.609,3.850)(0.615,3.851)(0.621,3.838)(0.626,3.800)(0.632,3.730)(0.637,3.628)(0.643,3.495)(0.649,3.339)(0.654,3.168)(0.660,2.993)(0.665,2.819)(0.671,2.653)(0.677,2.495)(0.682,2.346)(0.688,2.204)(0.693,2.066)(0.699,1.932)(0.704,1.799)(0.710,1.668)(0.716,1.539)(0.721,1.413)(0.727,1.291)(0.732,1.176)(0.738,1.071)(0.744,0.977)(0.749,0.895)(0.755,0.826)(0.760,0.770)(0.766,0.724)(0.772,0.689)(0.777,0.664)(0.783,0.646)(0.788,0.635)(0.794,0.627)(0.800,0.619)(0.805,0.606)(0.811,0.583)(0.816,0.548)(0.822,0.500)(0.828,0.440)(0.833,0.373)(0.839,0.302)(0.844,0.234)(0.850,0.173)(0.856,0.121)(0.861,0.081)(0.867,0.051)(0.872,0.031)(0.878,0.017)(0.883,0.009)(0.889,0.005)(0.895,0.002)(0.900,0.001)(0.900,0.000)}}
 \def\densitymoveOUAT{\density{move OUAT}{21}{(0.128,0.000)(0.128,0.009)(0.131,0.026)(0.134,0.067)(0.138,0.154)(0.141,0.316)(0.145,0.583)(0.148,0.969)(0.151,1.454)(0.155,1.982)(0.158,2.477)(0.162,2.873)(0.165,3.146)(0.168,3.322)(0.172,3.460)(0.175,3.620)(0.179,3.838)(0.182,4.126)(0.185,4.476)(0.189,4.878)(0.192,5.325)(0.196,5.803)(0.199,6.285)(0.202,6.727)(0.206,7.095)(0.209,7.374)(0.213,7.586)(0.216,7.769)(0.219,7.961)(0.223,8.176)(0.226,8.406)(0.230,8.630)(0.233,8.831)(0.236,8.995)(0.240,9.117)(0.243,9.191)(0.246,9.211)(0.250,9.168)(0.253,9.046)(0.257,8.833)(0.260,8.518)(0.263,8.105)(0.267,7.612)(0.270,7.069)(0.274,6.514)(0.277,5.978)(0.280,5.483)(0.284,5.035)(0.287,4.629)(0.291,4.245)(0.294,3.858)(0.297,3.446)(0.301,3.000)(0.304,2.531)(0.308,2.072)(0.311,1.662)(0.314,1.333)(0.318,1.098)(0.321,0.948)(0.325,0.859)(0.328,0.801)(0.331,0.746)(0.335,0.678)(0.338,0.597)(0.342,0.509)(0.345,0.427)(0.348,0.359)(0.352,0.309)(0.355,0.272)(0.358,0.243)(0.362,0.215)(0.365,0.188)(0.369,0.161)(0.372,0.139)(0.375,0.126)(0.379,0.126)(0.382,0.139)(0.386,0.163)(0.389,0.192)(0.392,0.218)(0.396,0.237)(0.399,0.244)(0.403,0.243)(0.406,0.236)(0.409,0.230)(0.413,0.228)(0.416,0.229)(0.420,0.231)(0.423,0.228)(0.426,0.215)(0.430,0.192)(0.433,0.160)(0.437,0.123)(0.440,0.087)(0.443,0.056)(0.447,0.033)(0.450,0.018)(0.454,0.009)(0.457,0.004)(0.460,0.002)(0.464,0.001)(0.464,0.000)}{(0.207,0.000)(0.207,0.012)(0.213,0.026)(0.219,0.052)(0.225,0.099)(0.230,0.175)(0.236,0.290)(0.242,0.448)(0.248,0.648)(0.254,0.880)(0.260,1.124)(0.266,1.354)(0.271,1.548)(0.277,1.688)(0.283,1.774)(0.289,1.814)(0.295,1.828)(0.301,1.835)(0.307,1.853)(0.313,1.892)(0.318,1.955)(0.324,2.043)(0.330,2.154)(0.336,2.283)(0.342,2.425)(0.348,2.574)(0.354,2.721)(0.360,2.856)(0.365,2.974)(0.371,3.070)(0.377,3.147)(0.383,3.212)(0.389,3.273)(0.395,3.335)(0.401,3.402)(0.407,3.473)(0.412,3.545)(0.418,3.613)(0.424,3.672)(0.430,3.721)(0.436,3.759)(0.442,3.785)(0.448,3.800)(0.453,3.807)(0.459,3.805)(0.465,3.795)(0.471,3.777)(0.477,3.750)(0.483,3.709)(0.489,3.653)(0.495,3.578)(0.500,3.483)(0.506,3.369)(0.512,3.237)(0.518,3.090)(0.524,2.934)(0.530,2.773)(0.536,2.611)(0.542,2.452)(0.547,2.298)(0.553,2.150)(0.559,2.007)(0.565,1.869)(0.571,1.734)(0.577,1.600)(0.583,1.466)(0.589,1.333)(0.594,1.204)(0.600,1.082)(0.606,0.971)(0.612,0.876)(0.618,0.796)(0.624,0.730)(0.630,0.674)(0.636,0.622)(0.641,0.570)(0.647,0.517)(0.653,0.463)(0.659,0.411)(0.665,0.364)(0.671,0.325)(0.677,0.295)(0.682,0.275)(0.688,0.261)(0.694,0.251)(0.700,0.239)(0.706,0.224)(0.712,0.204)(0.718,0.179)(0.724,0.150)(0.729,0.120)(0.735,0.091)(0.741,0.065)(0.747,0.044)(0.753,0.028)(0.759,0.017)(0.765,0.010)(0.771,0.005)(0.776,0.003)(0.782,0.001)(0.788,0.001)(0.788,0.000)}}
 \def\densitycombatives{\density{combatives}{13}{(0.074,0.000)(0.074,0.002)(0.078,0.009)(0.081,0.031)(0.085,0.087)(0.089,0.203)(0.093,0.402)(0.097,0.678)(0.101,0.994)(0.105,1.295)(0.109,1.558)(0.113,1.800)(0.117,2.064)(0.121,2.370)(0.125,2.718)(0.129,3.103)(0.133,3.538)(0.137,4.043)(0.140,4.609)(0.144,5.203)(0.148,5.779)(0.152,6.303)(0.156,6.758)(0.160,7.154)(0.164,7.540)(0.168,7.978)(0.172,8.496)(0.176,9.058)(0.180,9.579)(0.184,9.990)(0.188,10.264)(0.192,10.407)(0.195,10.417)(0.199,10.262)(0.203,9.906)(0.207,9.369)(0.211,8.734)(0.215,8.101)(0.219,7.513)(0.223,6.951)(0.227,6.381)(0.231,5.794)(0.235,5.200)(0.239,4.596)(0.243,3.961)(0.247,3.285)(0.251,2.597)(0.254,1.959)(0.258,1.423)(0.262,1.015)(0.266,0.737)(0.270,0.564)(0.274,0.465)(0.278,0.406)(0.282,0.365)(0.286,0.336)(0.290,0.319)(0.294,0.316)(0.298,0.327)(0.302,0.346)(0.306,0.365)(0.309,0.371)(0.313,0.362)(0.317,0.342)(0.321,0.323)(0.325,0.318)(0.329,0.328)(0.333,0.350)(0.337,0.375)(0.341,0.392)(0.345,0.386)(0.349,0.350)(0.353,0.290)(0.357,0.223)(0.361,0.169)(0.365,0.143)(0.368,0.148)(0.372,0.181)(0.376,0.231)(0.380,0.283)(0.384,0.319)(0.388,0.325)(0.392,0.299)(0.396,0.251)(0.400,0.199)(0.404,0.158)(0.408,0.137)(0.412,0.141)(0.416,0.164)(0.420,0.191)(0.424,0.204)(0.427,0.192)(0.431,0.160)(0.435,0.119)(0.439,0.080)(0.443,0.048)(0.447,0.026)(0.451,0.013)(0.455,0.005)(0.459,0.002)(0.463,0.001)(0.463,0.000)}{(0.103,0.000)(0.103,0.007)(0.109,0.014)(0.116,0.029)(0.122,0.056)(0.129,0.102)(0.136,0.172)(0.142,0.275)(0.149,0.411)(0.155,0.580)(0.162,0.773)(0.168,0.973)(0.175,1.164)(0.181,1.327)(0.188,1.451)(0.194,1.535)(0.201,1.584)(0.207,1.611)(0.214,1.632)(0.220,1.658)(0.227,1.699)(0.234,1.756)(0.240,1.829)(0.247,1.914)(0.253,2.006)(0.260,2.100)(0.266,2.194)(0.273,2.285)(0.279,2.373)(0.286,2.459)(0.292,2.547)(0.299,2.640)(0.305,2.741)(0.312,2.852)(0.318,2.972)(0.325,3.095)(0.332,3.214)(0.338,3.324)(0.345,3.417)(0.351,3.491)(0.358,3.545)(0.364,3.581)(0.371,3.600)(0.377,3.606)(0.384,3.597)(0.390,3.572)(0.397,3.528)(0.403,3.462)(0.410,3.372)(0.417,3.262)(0.423,3.136)(0.430,3.003)(0.436,2.870)(0.443,2.744)(0.449,2.626)(0.456,2.516)(0.462,2.410)(0.469,2.305)(0.475,2.197)(0.482,2.087)(0.488,1.976)(0.495,1.864)(0.501,1.756)(0.508,1.651)(0.515,1.550)(0.521,1.450)(0.528,1.350)(0.534,1.250)(0.541,1.150)(0.547,1.052)(0.554,0.958)(0.560,0.872)(0.567,0.795)(0.573,0.730)(0.580,0.676)(0.586,0.633)(0.593,0.599)(0.599,0.571)(0.606,0.548)(0.613,0.526)(0.619,0.504)(0.626,0.479)(0.632,0.451)(0.639,0.420)(0.645,0.385)(0.652,0.346)(0.658,0.304)(0.665,0.260)(0.671,0.216)(0.678,0.172)(0.684,0.132)(0.691,0.097)(0.697,0.068)(0.704,0.045)(0.711,0.028)(0.717,0.017)(0.724,0.010)(0.730,0.005)(0.737,0.003)(0.743,0.001)(0.750,0.001)(0.750,0.000)}}
 \def\densitycasualtyevacuation{\density{casualty evac}{6}{(0.002,0.000)(0.002,0.012)(0.006,0.038)(0.011,0.110)(0.015,0.272)(0.019,0.581)(0.024,1.079)(0.028,1.759)(0.032,2.538)(0.037,3.293)(0.041,3.909)(0.045,4.341)(0.049,4.609)(0.054,4.772)(0.058,4.904)(0.062,5.073)(0.067,5.325)(0.071,5.671)(0.075,6.089)(0.080,6.546)(0.084,7.009)(0.088,7.441)(0.092,7.797)(0.097,8.037)(0.101,8.161)(0.105,8.208)(0.110,8.224)(0.114,8.225)(0.118,8.187)(0.123,8.074)(0.127,7.864)(0.131,7.570)(0.135,7.219)(0.140,6.843)(0.144,6.461)(0.148,6.075)(0.153,5.675)(0.157,5.259)(0.161,4.843)(0.166,4.447)(0.170,4.069)(0.174,3.674)(0.179,3.219)(0.183,2.686)(0.187,2.108)(0.191,1.550)(0.196,1.079)(0.200,0.733)(0.204,0.510)(0.209,0.380)(0.213,0.308)(0.217,0.267)(0.222,0.245)(0.226,0.235)(0.230,0.236)(0.234,0.241)(0.239,0.244)(0.243,0.242)(0.247,0.234)(0.252,0.226)(0.256,0.223)(0.260,0.228)(0.265,0.238)(0.269,0.249)(0.273,0.254)(0.278,0.249)(0.282,0.233)(0.286,0.211)(0.290,0.194)(0.295,0.191)(0.299,0.206)(0.303,0.234)(0.308,0.263)(0.312,0.281)(0.316,0.282)(0.321,0.270)(0.325,0.254)(0.329,0.242)(0.333,0.237)(0.338,0.234)(0.342,0.226)(0.346,0.212)(0.351,0.195)(0.355,0.182)(0.359,0.180)(0.364,0.188)(0.368,0.199)(0.372,0.205)(0.377,0.202)(0.381,0.191)(0.385,0.175)(0.389,0.157)(0.394,0.136)(0.398,0.110)(0.402,0.082)(0.407,0.054)(0.411,0.031)(0.415,0.016)(0.420,0.007)(0.424,0.003)(0.428,0.001)(0.428,0.000)}{(-0.001,0.000)(-0.001,0.007)(0.004,0.016)(0.009,0.032)(0.014,0.060)(0.019,0.108)(0.025,0.182)(0.030,0.290)(0.035,0.434)(0.040,0.614)(0.045,0.822)(0.050,1.040)(0.055,1.249)(0.060,1.428)(0.065,1.561)(0.070,1.643)(0.075,1.680)(0.081,1.686)(0.086,1.681)(0.091,1.684)(0.096,1.707)(0.101,1.757)(0.106,1.832)(0.111,1.926)(0.116,2.032)(0.121,2.145)(0.126,2.264)(0.131,2.390)(0.137,2.524)(0.142,2.666)(0.147,2.818)(0.152,2.976)(0.157,3.139)(0.162,3.304)(0.167,3.469)(0.172,3.631)(0.177,3.786)(0.182,3.927)(0.187,4.047)(0.193,4.142)(0.198,4.209)(0.203,4.251)(0.208,4.271)(0.213,4.278)(0.218,4.277)(0.223,4.272)(0.228,4.264)(0.233,4.250)(0.238,4.225)(0.243,4.184)(0.248,4.125)(0.254,4.048)(0.259,3.954)(0.264,3.846)(0.269,3.729)(0.274,3.607)(0.279,3.483)(0.284,3.359)(0.289,3.237)(0.294,3.115)(0.299,2.992)(0.304,2.866)(0.310,2.737)(0.315,2.605)(0.320,2.474)(0.325,2.347)(0.330,2.227)(0.335,2.117)(0.340,2.014)(0.345,1.917)(0.350,1.819)(0.355,1.718)(0.360,1.611)(0.366,1.496)(0.371,1.377)(0.376,1.259)(0.381,1.149)(0.386,1.051)(0.391,0.970)(0.396,0.907)(0.401,0.859)(0.406,0.822)(0.411,0.789)(0.416,0.754)(0.422,0.711)(0.427,0.657)(0.432,0.591)(0.437,0.514)(0.442,0.431)(0.447,0.348)(0.452,0.268)(0.457,0.198)(0.462,0.139)(0.467,0.093)(0.472,0.059)(0.478,0.036)(0.483,0.020)(0.488,0.011)(0.493,0.006)(0.498,0.003)(0.503,0.001)(0.503,0.000)}}
 \def\densitytotal{\density{composite}{17}{(0.088,0.000)(0.088,0.012)(0.092,0.031)(0.095,0.072)(0.099,0.155)(0.103,0.304)(0.107,0.548)(0.111,0.908)(0.115,1.386)(0.119,1.957)(0.123,2.571)(0.127,3.161)(0.131,3.665)(0.135,4.043)(0.139,4.288)(0.143,4.420)(0.147,4.478)(0.151,4.512)(0.155,4.564)(0.159,4.665)(0.163,4.829)(0.167,5.048)(0.171,5.297)(0.175,5.548)(0.179,5.774)(0.183,5.962)(0.187,6.115)(0.191,6.246)(0.195,6.367)(0.199,6.484)(0.203,6.598)(0.207,6.703)(0.211,6.790)(0.215,6.852)(0.219,6.883)(0.223,6.877)(0.227,6.831)(0.231,6.740)(0.235,6.600)(0.239,6.407)(0.243,6.165)(0.247,5.888)(0.251,5.597)(0.255,5.315)(0.259,5.057)(0.262,4.825)(0.266,4.601)(0.270,4.360)(0.274,4.079)(0.278,3.750)(0.282,3.381)(0.286,2.994)(0.290,2.612)(0.294,2.256)(0.298,1.935)(0.302,1.650)(0.306,1.398)(0.310,1.177)(0.314,0.991)(0.318,0.840)(0.322,0.728)(0.326,0.650)(0.330,0.601)(0.334,0.570)(0.338,0.550)(0.342,0.533)(0.346,0.514)(0.350,0.491)(0.354,0.463)(0.358,0.430)(0.362,0.395)(0.366,0.363)(0.370,0.333)(0.374,0.307)(0.378,0.286)(0.382,0.268)(0.386,0.252)(0.390,0.240)(0.394,0.231)(0.398,0.224)(0.402,0.223)(0.406,0.227)(0.410,0.239)(0.414,0.257)(0.418,0.279)(0.422,0.300)(0.426,0.318)(0.429,0.325)(0.433,0.319)(0.437,0.299)(0.441,0.264)(0.445,0.219)(0.449,0.169)(0.453,0.122)(0.457,0.081)(0.461,0.049)(0.465,0.028)(0.469,0.014)(0.473,0.007)(0.477,0.003)(0.481,0.001)(0.481,0.000)}{(0.168,0.000)(0.168,0.015)(0.171,0.031)(0.174,0.059)(0.178,0.105)(0.181,0.180)(0.185,0.292)(0.188,0.450)(0.191,0.661)(0.195,0.925)(0.198,1.234)(0.202,1.572)(0.205,1.918)(0.208,2.246)(0.212,2.531)(0.215,2.759)(0.219,2.924)(0.222,3.029)(0.226,3.087)(0.229,3.113)(0.232,3.123)(0.236,3.129)(0.239,3.141)(0.243,3.162)(0.246,3.196)(0.249,3.245)(0.253,3.313)(0.256,3.402)(0.260,3.513)(0.263,3.646)(0.267,3.796)(0.270,3.954)(0.273,4.112)(0.277,4.259)(0.280,4.389)(0.284,4.497)(0.287,4.585)(0.290,4.657)(0.294,4.717)(0.297,4.774)(0.301,4.830)(0.304,4.887)(0.307,4.947)(0.311,5.005)(0.314,5.061)(0.318,5.111)(0.321,5.154)(0.325,5.188)(0.328,5.213)(0.331,5.228)(0.335,5.233)(0.338,5.230)(0.342,5.220)(0.345,5.204)(0.348,5.181)(0.352,5.151)(0.355,5.109)(0.359,5.053)(0.362,4.982)(0.365,4.893)(0.369,4.790)(0.372,4.678)(0.376,4.564)(0.379,4.455)(0.383,4.356)(0.386,4.266)(0.389,4.183)(0.393,4.099)(0.396,4.007)(0.400,3.898)(0.403,3.767)(0.406,3.614)(0.410,3.443)(0.413,3.261)(0.417,3.076)(0.420,2.895)(0.424,2.721)(0.427,2.555)(0.430,2.395)(0.434,2.237)(0.437,2.077)(0.441,1.914)(0.444,1.745)(0.447,1.571)(0.451,1.393)(0.454,1.213)(0.458,1.036)(0.461,0.863)(0.464,0.701)(0.468,0.552)(0.471,0.421)(0.475,0.310)(0.478,0.219)(0.482,0.149)(0.485,0.097)(0.488,0.060)(0.492,0.036)(0.495,0.020)(0.499,0.011)(0.502,0.006)(0.505,0.003)(0.505,0.000)}}

\begin{figure*}\centering
\definecolor{bblue}{rgb}{.2,.7,1}
\newcommand\density[4]{
\begin{tikzpicture}
\begin{axis}[width=.21\textwidth,height=.15\textwidth,scale only axis,
 axis line style={draw=none},xtick=\empty, clip=false, 
ytick=\empty]
\draw [white] (rel axis cs:0,0) rectangle (rel axis cs:1,1);
\addplot[draw=none,fill=bblue, fill opacity=1] coordinates{#3};
\addplot[draw=none,fill=red, fill opacity=0.5] coordinates{#4};
\node[align=center,below=0.25em] at  (rel axis cs:0.5,0) {\smash{#1 (#2:1)}};
\end{axis}\end{tikzpicture}}
\densityhastyfighting\hspace{-1em}%
\densitymoveOUAT\hspace{-1em}%
\densitycombatives\hspace{-1em}%
\densitycasualtyevacuation\hspace{-1em}%
\densitytotal\\
\renewcommand\density[4]{
\begin{tikzpicture}
\begin{axis}[width=.26\textwidth,height=.15\textwidth,scale only axis,
 axis line style={draw=none},xtick=\empty, clip=false, 
ytick=\empty]
\draw [white] (rel axis cs:0,0) rectangle (rel axis cs:1,1);
\addplot[draw=none,fill=bblue, fill opacity=1] coordinates{#3};
\addplot[draw=none,fill=red, fill opacity=0.5] coordinates{#4};
\node[align=center,below=0.25em] at  (rel axis cs:0.5,0) {\smash{#1 (#2:1)}};
\end{axis}\end{tikzpicture}}
\densitysleddrag\hspace{-1.5em}\densityrun\hspace{-1.5em}\densitydeadlift\hspace{-1.5em}\densitysledpush\\
\densitypushups\hspace{-1.5em}\densitypthrow\hspace{-1.5em}\densitysitups\hspace{-1.5em}\densitylegtuck
\caption[no caption]{Density plots of Fort Riley fighting-load \WTST\ vignettes and several test events for men \tikz{\fill[bblue, fill opacity=1] (0,0) rectangle (0.8em,0.8em);} and women \tikz{\fill[red, fill opacity=0.5] (0,0) rectangle (0.8em,0.8em);}. 
The smaller the relative overlap
\tikz{\fill[bblue, fill opacity=1] (0,0) rectangle (0.8em,0.8em);\fill[red, fill opacity=0.5] (0,0) rectangle (0.8em,0.8em);}, the greater the gender bias in each event. The odds that a male soldier chosen at random outperforms a female soldier chosen at random are listed in the parentheses.}
\label{Fig:4}
\end{figure*}
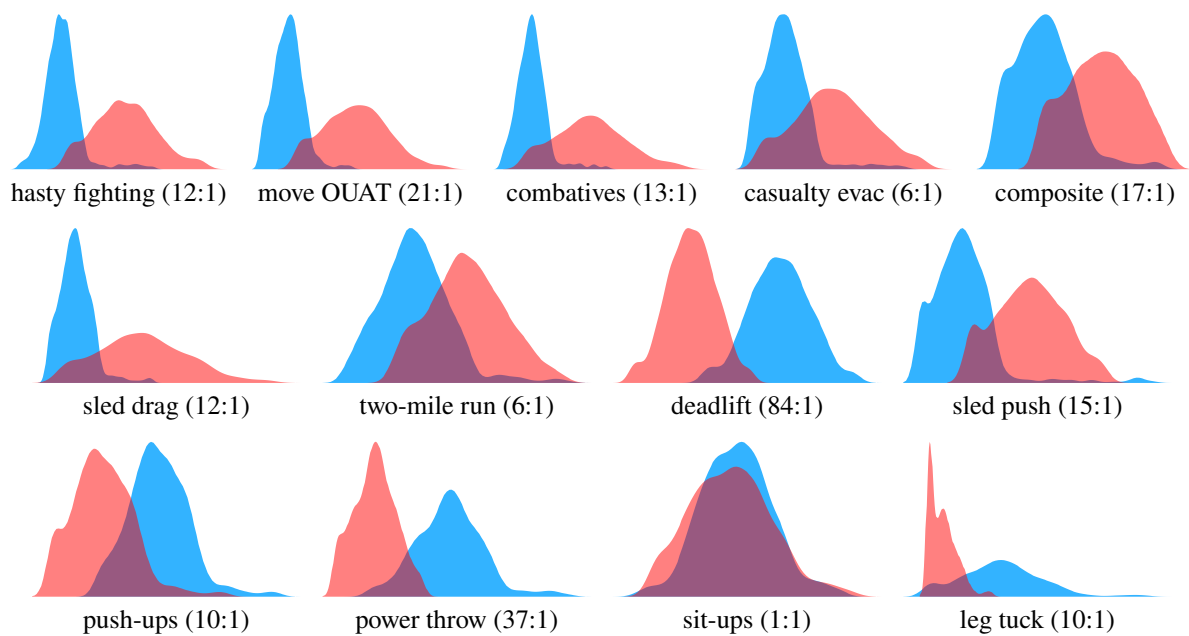

Without having access to the original data set, it is impossible to fully analyze the gender bias that is built into the BSPRRS model.
However, we can make a fair estimate of the gender by computing the effect size based on differences in the the mean scores of men and women for each test event.%
\endnote{
In making effect size readily understood by non-statisticians, researchers McGraw and Wong (1992) have suggested using a Common Language Effect Size (CLES), which gives the probability that a score sampled at random from one distribution would be greater than a score sampled at random from another distribution. This probability is the standard normal cumulative distribution function evaluated at $d/\sqrt{2}$ where $d$ is Cohen's standardized mean difference $d$.  Cohen's $d$ is the difference of the two means divided by the pooled standard deviation: $d = (\overline{x}_1 - \overline{x}_2)/s$ where $s = \sqrt{(n_1s_1^2 + n_2s_2^2)/(n_1+n_2)}$.  This makes a strong assumption on normality of the data, the same assumptions that the BSPRRS researchers make. An alternative approach that doesn't assume normality is by sampling from the distribution. The BSPRRS final report lists scores at several percentiles for each test event: 0, 5, 10, 25, 50, 75, 90, 95, 100. We can use a piecewise cubic Hermite interpolant (or a piecewise linear inerpolant) through the points to reconstruct the distributions and then use the Monte Carlo method to draw samples of female and male soldiers for comparison. To make CLES even more readily understood, we can reinterpret it as an odds $p:1-p$, and then normalizing the ratio $p/(1-p):1$. }
One way to interpret the effect size is as the odds that a male soldier chosen at random would outperform a female soldier chosen at random on a given test event.
For example, the odds of a male soldier outperforming a female soldier on the two-mile run are 6:1. 
The odds of a male soldier outperforming a female soldier in push-ups are 10:1, and in sit-ups the odds are an even 1:1. 
The deadlift is the most gender-biased test event of the 23 candidate events with an odds of 84:1.
The gender bias for Fort Riley test events  is shown in Figure~\ref{Fig:4}.\endnote{The density plots are derived by Monte Carlo sampling ($n=1000$) the distribution reconstructed using a piecewise cubic Hermite interpolant (pchip) of the percentiles listed in the BSPRRS final report. The curve is smoothed using a Gaussian kernel.}

The researchers state that  \textQT{while predictive validity was crucial, it was equally as important to the Army to produce a test that assessed all components of fitness. A multi-component physical assessment was essential to transform physical readiness training and reduce musculoskeletal injuries.} Army leadership was concerned that these seven test events  \textQT{did not represent all components of physical fitness and therefore would not drive a comprehensive change in physical readiness training to increase combat lethality and potentially reduce musculoskeletal injuries.} Army leadership was also concerned about the lack of anaerobic endurance and core strength test events. So, the kettlebell squat was removed as a test event and the 300-yard shuttle run and the leg tuck were ``forced into the model,'' although neither are significant predictors of the \WTST\ composite times.   Table~\ref{table:4} gives  the model coefficients and Figure~\ref{Fig:5} shows the relative importance of each event in the eight-test-event predictor model. 

Under the new model, the sled drag was nine times as important as either the shuttle run or the leg tuck. And both the leg tuck and the shuttle run each contributed to less than 5 percent of the models predictive variability.\endnote{By taking the dot product of the coefficients and the means added to the constant term, we get a total of 831 seconds, which agrees closely with the average composite \WTST\ time of 841 seconds. Taking the dot product of the absolute values of the coefficients and the standard deviations, we get a total of 235 seconds, which agrees closely with the average composite \WTST\ standard deviation of 234 seconds.}

During testing at Fort Riley, almost fifty percent of female soldiers and five percent of male soldiers were unable to complete even one leg tuck. But, because leg tucks were determined to have such little predictive weight in this model, each complete leg tuck subtracts only two seconds from the predicted \WTST\ composite time. In other words, failure to do one leg tuck is equivalent to adding five seconds onto their two-mile run time or doing two fewer push-ups.

\sparagraph{Fort Benning}
Model cross-validation is an absolutely essential step in developing a predictive model. Cross-validation tests a model's ability to make predictions given new information about new soldiers and not just  the 339 soldiers whose data was already used to make the model. It helps to identify common problems like overfitting and selection bias. And it helps build trust that the model can be generalized across the whole of the Army. To do this, new data (fitness test event scores from a different representative group of soldiers) needs to be collected as input data to the predictor model. The output of the model (their predicted composite \WTST\ times) is then compared with the actual collected data (their actual composite \WTST\ times), and the mean squared error is calculated. Ideally, the mean squared error of the validation data should be close to the mean squared error of the training data.
 
But, the Army never validated the model. The researchers conducted a follow-up event  using 136 male and 16 female soldiers\endnote{It's concerning that the Army used only 16 women to represent fitness standards that impact 74,000 women. Women accounted for less than 11 percent of the study participants, less than even the 15 percent composition of the Army.} (all volunteers) at Fort Benning with what appears to be the intent of validating the Fort Riley eight-test-event model. Data was collected on \WTST\ completion times and the eight predictive test event scores (the test events that eventually went into building the ACFT). However, instead of using the Fort Benning event to cross-validate the Fort Riley model, the researchers did another ``full model regression analysis utilizing the empirical raw scores.'' It's curious why the researchers, instead of cross-validating the model, decided to create an entirely new model instead. Did the Fort Riley model fail validation using Fort Benning data? Were the test events at Fort Benning so different from test events at Fort Riley as  to make comparison of data impossible?  By starting over, are the researchers acknowledging a fundamental error in the model design?   Table~\ref{table:5} shows the new linear regression coefficients along with data from Fort Benning.\endnote{The data from Fort Benning included in the final report (Table 17) contains numerous errors that makes it difficult to analyze.}

There are several noticeable differences between the Fort Riley and the Fort Benning data. The mean times of the sled drag and the sled push at Fort Riley are 18.3 and 8.8 seconds respectively. At Fort Benning the times were 67 seconds and 33 seconds—about 3.7 times longer. Why was the distance changed for these two test events? Additionally, the mean two-mile run time increased from 885 seconds to 1011 seconds, a change of over a standard deviation. But, the most striking change between Fort Riley and Fort Benning is the model. We can see the difference by simply comparing coefficients. For some unknown reason it appears that rather than using the composite \WTST\ time for all four vignettes, the researchers used only the time for the first, hasty-fighting-position vignette to develop their new predictive linear regression model. To see this, we can apply the model to the mean test event scores to give a mean predicted time of 268 seconds, which agrees closely with the average \WTST\ hasty-fighting-position mean of 262 seconds, and is quite different from the composite time of 606 seconds. For further confirmation we can also compare the standard deviations in the test event scores and the \WTST\ vignette times.\endnote{The standard deviation predicted by the model should be close to the weighted $L^1$ or $L^2$ norm (depending on the multicollinearity of the variables) of the standard deviation divided by the square root of $R^2$.} The model predicts a standard deviation of about 44 seconds, which agrees moderately well with the average \WTST\ hasty-fighting-position standard deviation of 67 seconds, compared to a 202-second standard deviation across all four vignettes.
 
In the Fort Benning model, the two-mile run accounts for almost 70 percent of the variance. The leg tuck and the push-up on the other hand account for less than one percent. Furthermore, the coefficient of the leg tuck is positive, saying that it is anti-correlated with improved \WTST\ scores. The leg tuck and push-up should not be used as predictors in the model (which applies to the hasty-fighting-position vignette).

Following the conclusion of the study, Army senior leaders asked that the sled drag, sled push, and shuttle run be combined into one test event, the sprint-drag-carry (SDC), to reduce administration time, equipment costs, and the total number of test events.
The SDC is composed of a 50-m sprint, a 50-m sled drag, a 50-m  sideways lateral dash, a 50-m farmers carry, and finally an additional 50-m sprint. There are clear concerns about making such a substitution—it is not the same test event! In an attempt to incorporate this post hoc change into the Fort Benning model, the researchers consolidated the  sled drag, sled push, and shuttle run variables into one composite variable based on the standardized values of these three test events.\endnote{We can compose a composite mean score by summing the test event means multiplied by their respective standard deviations and dividing the quantity by the square root of the sum of the squares of the standard deviations. We can compose a composite standard deviation by taking the square root of the sum of the squares of the standard deviations.}
Table~\ref{table:6} shows the new linear regression coefficients along with data from Fort Benning.\endnote{By taking the dot product of the coefficients and the means added to the constant term, we get a total of 248 seconds, which agrees closely with the average \WTST\ hasty-fighting-position vignette mean of 268 seconds. Taking the dot product of the absolute values of the coefficients and the standard deviations, we get a total of 88 seconds, which agrees moderately well with the average \WTST\ hasty-fighting-position vignette standard deviation of 67 seconds.} Figure~\ref{Fig:6} shows the relative importance of these test events in the model.


\begin{table*}\centering
\begin{tabularx}{1\linewidth}{>{\rowmac}l *{9}{>{\rowmac}Y}<{\clearrow}}
\toprule
event & constant & \makebox[0pt][c]{sled drag} & run & deadlift & \makebox[0pt][c]{sled push} & \makebox[0pt][c]{push-ups} & p.~throw & \makebox[0pt][c]{shuttle run} & leg tuck \\\midrule
coefficient & 6.72 & 0.049 & 0.28 & $-$0.012 & 0.016 & $-$0.003 & $-$0.099 & $-$0.27 & 0.02 \\
mean & \NA & 67.0 & 1011 & 229.5 & 33.1 & 52.5 & 20.6 & 70.5 & 9.2 \\
SD & \NA & 48.3 & 122 & 46.2 & 27.0 & 14.3 & 5.7 & 6.5 & 5.8 \\\midrule
\end{tabularx}
\caption{Regression coefficients and importance for Fort Benning eight-test-event predictor model.}
\label{table:5}
\end{table*}
\begin{table*}\centering
\begin{tabularx}{0.78\linewidth}{>{\rowmac}l *{7}{>{\rowmac}Y}<{\clearrow}}
\toprule
event &  constant & SDC & run & deadlift & \makebox[0pt][c]{push-ups} & \makebox[0pt][c]{p.~throw} &\makebox[0pt][c]{leg tuck}\\\midrule
coefficient & 12.21 & 1.20 & 0.16 & $-$0.015 & 0.021 & $-$0.082 & 0.031\\
mean & \NA &{\itshape 82.4} & 1011 & 229.5 & 52.5 & 20.6 & 9.2 \\
SD & \NA & {\itshape 55.8} & 122 & 46.2 & 14.3 & 5.7 & 5.8 \\\midrule
\end{tabularx}
\caption{Regression coefficients and importance for final Fort Benning six-test-event predictor model.}
\label{table:6}
\end{table*}
 
The final report does not provide data on the amalgamated SDC test event outside of stating that the data was computed using the z-scores of the sled push, sled drag, and shuttle run. 
The mean and standard deviations listed in Table~\ref{table:6} for the SDC are likewise derived  using z-scores to combine the point estimates from these three events.
\endnote{The composite mean $\mu_\ast$ and standard deviation $\sigma_\ast$ are computed using weighted z-scores
$\mu_{\ast}= (\sigma_1 \mu_1 + \sigma_2 \mu_2 + \sigma_3 \mu_3)/\sigma_\ast$ where $\sigma_{\ast}=\sqrt{\sigma^2_1  + \sigma^2_2  + \sigma^2_3}$ and  $\mu_i$ and $\sigma_i$ are the mean and standard deviations of the sled push, sled drag, and shuttle run.}
The relative importance of these test events in the adjusted  model are shown in Figure~\ref{Fig:7}.
The score is not representative of a real sprint-drag-carry test event. The  sprint-drag-carry and two-mile run now dominate the six-test-event model as predictors. The other four events are all less than one percent. Moreover, both the leg tuck and push-up have positive coefficients.
Saying that leg tucks (or push-ups) have any real relevance as predictors in the model is a bit like giving credit to an airline passenger for a safe landing, simply because he was included on the flight manifest along with the pilot and other crew members.
So, there it is---the model that the Army has claimed as being 80 percent predictive.

\sparagraph{The Army Combat Fitness Test}
In moving from a predictive model to a physical fitness evaluation framework, there are several things that should be remembered. A predictive linear regression model does not have minimum scores, only a cumulative score. Being faster or stronger in one category can offset being slower or weaker in another. The test events in a linear model are each weighted differently and determined by the relationship of the predictor variables to the outcome variables. The two-mile run and the  sprint-drag-carry  are the most predictive events, while leg tucks and push-ups are the least. In the Fort Riley model leg tucks have a relative importance of four percent and in the flawed Fort Benning model leg tucks are less than one percent, while the sprint-drag-carry and two-mile run are together well over 70 percent. Women were vastly underrepresented in the model training data. The model that the Army erroneously claims as having an 80 percent ability to predict WTBD/CST for every soldier was developed using a mere 16 women. Moreover, the model was developed using volunteers who are not representative of the Army as a whole. In the Army's attempt to construct a gender neutral-test, they eliminated sit-ups, the one test that was truly gender neutral. Half of all women in the training couldn't do a leg tuck.



The ACFT consists of six  timed events, each based on tasks that a soldier might encounter in training or combat.\autocite{frost_2018__video}
The Army has claimed that the deadlift replicates a ``litter carry or the movement of ammunition and supplies.''
The power throw replicates ``the movement required to assist a buddy over an obstacle or the power required to leap across a ditch.''
The sprint-drag-carry replicates  ``moving a casualty to safety moving supplies or moving under fire.''
The push-up replicates  ``hand and arm movements required in combatives or repetitive loading of ammunition and supplies.''
The leg tuck replicates  ``climbing up and over walls obstacles or exiting disabled vehicles.''
And the two-mile run replicates ``movements to and from contact.''
Every soldier must pass each test event at a minimum gold, gray, or black standard depending on his or her military occupation. See Table~\ref{table:7}. 
Table~\ref{table:8} shows the percentage of soldiers in Fort Riley trials achieving at gold, gray, or black standard scores for male and female soldiers.
\endnote{Fort Benning data is not broken down by by gender in the BSPRRS final report, but overall it agrees with the Fort Riley data.}%
\textsuperscript{,}\endnote{Researchers tested participants using a five-repetition-maximum deadlift and converted then to the scores to an equivalent one-repetition-maximum deadlift using the Wathen formula:
$
 w_{1} = 100 w_r / (48.8 + 53.8\exp(-0.075 r))
$
where $w_r$ is the weight lifted for $r$ reps. The ACFT uses a three-repetition-maximum deadlift. For the purposes of the table, ACFT standards  have been adjusted to match the one-repetition-maximum scores listed in the BSPRRS final report. Because of changes from five-repetition-maximum to three-repetition-maximum using one-repetition-maximum, the scores listed in the table may not be representative.}
The bottom row shows the percentage of soldiers reported by the Army during the first six months of ACFT implementation.\endnote{Information Paper, ``FY19 ACFT Test Battalions---Additional RFI Answers,'' July 9,2020.}

During six months of ACFT trials, with approximately fourteen thousand soldiers, 60 percent of all female soldiers and eight percent of all male soldiers were unable to do one leg tuck (the minimum number needed to pass.) See Table~\ref{table:8}. Shortly afterwards, due to the incredibly high leg-tuck failure rate, the Army allowed service members to temporarily substitute a two-minute plank in lieu of a leg tuck.\autocite{rempfer_2020} The plank was never included as one of the 23 test events examined during the development of the BSPRRS model. Exercise physiologists have noted that a plank and a leg tuck are very different assessments. A plank is an isometric assessment of core fitness. A leg tuck is a dynamic assessment of ``grip strength, shoulder adduction and flexion, elbow flexion, and trunk and hip flexion.''\autocite{acft_2018_manual} That the Army is substituting one test event with another test event of different muscle groups underscores the weak and arbitrary reasoning behind the leg tuck in the first place.\endnote{Personal communication with exercise physiology researchers Dr. Vince Tedjasaputra and Dr. Stewart Petersen.} A plank hardly replicates ``climbing up and over walls obstacles or exiting disabled vehicles.''\endnote{Many might argue that a leg tuck does npt replicate climbing up and over walls obstacles or exiting disabled vehicles either.}

\begin{table*}\centering
\begin{tabularx}{.95\linewidth}{>{\rowmac}l   c *{6}{>{\rowmac}Y}<{\clearrow}}
\toprule
standard & score & deadlift & \makebox[0pt][c]{power throw} & \makebox[0pt][c]{two-mile run}  & push-ups & leg tuck & SDC\\\midrule
maximum &  100 & 340 &12.5 & 13:30 & 60 & 20 &1:33 \\
black & 70 & 200 & 8.0 & 18:00 & 30 & 5 & 2:10 \\
gray & 65 & 180 & 6.5 & 19:00 &20 & 3 & 2:30 \\
gold &60 & 140 &4.5 & 21:00 & 10 & 1 & 3:00\\
minimum & 0& 80 & 3.3 & 22:48 & 0& 0 & 3:35\\
\midrule
\end{tabularx}
\caption{ACFT standards and scoring. Deadlifts are counted in pounds, power throw in meters, two-mile run and SDC in minutes, and push-ups and leg tuck in repetitions.}
\label{table:7}
\end{table*}

\definecolor{Yellow}{rgb}{1,1,1}
\newcolumntype{B}{>{\columncolor{Yellow}}Y}
\begin{table*}\centering
\newcommand\lrsplit[2]{\makebox[0pt][c]{\llap{#1} $\mid$ \rlap{#2}}}
\begin{tabularx}{.9\linewidth}{l *{4}{Y} B Y }
\toprule
standard & deadlift
 & power throw & \makebox[0pt][c]{two-mile run}  & push-ups & leg tuck & SDC\\\midrule
\setrow{\itshape}  maximum & \itshape\lrsplit{0}{0}& \itshape\lrsplit{96}{23} & \itshape\lrsplit{23}{1} & \itshape\lrsplit{69}{8} & \itshape\lrsplit{2}{0} &*\\
\setrow{\itshape}  black & \itshape\lrsplit{88}{3} & \itshape\lrsplit{100}{80} & \itshape\lrsplit{96}{81 } & \itshape\lrsplit{100}{80} & \itshape\lrsplit{74}{3} & *\\
\setrow{\itshape}  gray & \itshape\lrsplit{96}{11} & \itshape\lrsplit{100}{ 94} & \itshape\lrsplit{98}{97} & \itshape\lrsplit{100}{100} & \itshape\lrsplit{84}{19} & *\\
\setrow{\itshape}  gold &\itshape\lrsplit{100}{66} & \itshape\lrsplit{100}{100} & \itshape\lrsplit{100}{100} & \itshape\lrsplit{100}{100} & \itshape\lrsplit{92}{56} & *\\
\midrule
FY19 &\lrsplit{95}{92}&\lrsplit{96}{94}&\lrsplit{96}{84} &\lrsplit{96}{94} &\lrsplit{90}{17} &\lrsplit{82}{82}\\
FY20-Q1 &\lrsplit{98}{94}&\lrsplit{99}{96}&\lrsplit{97}{90} &\lrsplit{99}{96} &\lrsplit{92}{36} &\lrsplit{98}{88}\\
FY20-Q2 &\lrsplit{100}{99}& \lrsplit{100}{97}&\lrsplit{98}{91} &\lrsplit{100}{97} &\lrsplit{93}{46} &\lrsplit{99}{94} \\
\midrule
\end{tabularx}
\caption[no caption]{Percentage of soldiers in Fort Riley trials achieving at gold, gray, or black standard scores, split by male (left) and female soldiers (right). The bottom row is the percentage of soldiers reported by the Army.}
\label{table:8}
\end{table*}





Army Chief of Staff General Mark Milley said of the new ACFT ``This fitness test is hard. No one should be under any illusions about it.'' He's correct, unless he's referring to the roughly 40 percent of soldiers who are male and under twenty five.
Under the original APFT an 18-year old male soldier needed to do 42 push-ups to pass. But, under the new ACFT standards that same soldier needs to do only ten.\endnote{The ACFT uses hand-release push-ups whereas the AFPT uses a traditional push-up. A hand-release push-up requires  the chest to touch the ground and hands to be lifted up between repetitions. To some hand-release push-ups are more difficult and to others traditional push-ups are more difficult.} 
Put another way, an 18-year old under the ACFT standards needs to have the same upper-body muscular endurance as a 50-year-old female soldier and less than a 65-year-old male soldier, who needed to do an additional six push-ups under the old APFT.\endnote{The Army has not only reduced the minimum scores needed to pass, they also reduced the maximum scores. The ACFT maximum for push-ups is 60 repetitions, the same maximum expected of a 46-year-old male soldier under the old APFT standards.}
That 18-year-old soldier also doesn't need to have the same speed and aerobic endurance under the new ACFT as he did under the old APFT. The ACFT two-mile gold standard run time is 21 minutes.\endnote{The Army expects run scores to be slower due to the other strenuous activity of the ACFT.}
That 18-year-old male soldier under the ACFT is allowed an extra minute over what a 65-year-old man  needed under the APFT rules.
At the same time, the ACFT forces a 40-year-old woman   to run almost two minutes faster. 
The original APFT minimum standards were constructed using normative measures.
Thirty percent of 40-year-old female soldiers ran in over 21 minutes.\autocite{knapik1993army}
Even when training regularly, runners run slower with age.\autocite{devita2016relationships}
And many older soldiers have chronic injuries resulting from years of deployments and carrying heavy gear.\autocite{ryan_2020}
If the Army argues that every soldier must have the same levels of minimal fitness regardless of age, and that indeed, with a few months of training, any soldier can achieve gold standard, why does the Army set the maximum age to enlist for active duty at 34 years?

In April 2020 the Army tested a battery of basic trainees 
at Fort Sill, Oklahoma  using both the APFT and the ACFT. 
They found men had a 60 percent fail rate on the APFT but only a 3 percent fail rate on the ACFT. Women also had 60 percent fail rate on the APFT but  a 47 percent fail rate on the APFT.%
\endnote{From Update from the SMA (Sergeant Major of the Army) July 2020 to U.S.~Army Center for Initial Military Training: Trainees from the C/1-40 FA(BCT) took the  ACFT on April 20, 2020 followed by the APFT on April 24, 2020. The number of soldiers who passed or failed are as follows:\\
	\begin{tabularx}{1\linewidth}{lccccccc}\toprule
	 &male pass & male fail & female pass & female fail\\\midrule
	ACFT &98 & 3 & 50 & 45 \\
	APFT &  44 & 68 & 31 & 46 \\\midrule
\end{tabularx}} 
The Army has effectively made their fitness test of record 20 times easier for male recruits and 1.3 times easier for female recruits.
The fail rates on the APFT were largely due to the two-mile-run, and the fail rates on the ACFT were largely due to the leg tuck. In going from the APFT to the ACFT, the Army has made the two-mile-run much easier, especially for young male soldiers.  
At the same time, they added leg tucks that predominately adversely affect female soldiers. 







The 2015 NDAA requires that gender-neutral occupational standards first ``accurately predict performance of actual, regular, and recurring duties of a military occupation,'' and second ``are applied equitably to measure individual capabilities.'' RAND has offered the following two criteria for assessing where physical fitness tests are equitably applied: \textQT{Test validity should not differ among relevant subgroups (such as gender and race), and test scores should be unbiased (i.e., two people who receive the same test score should have the same likelihood of success on the job, regardless of subgroup).} The ACFT fails to meet either of these requirements for a valid, unbiased gender-neutral test.

Adverse impact, often referred to as unintentional discrimination, is any employment policy or practice that disproportionately and adversely affects one group of people of a protected class (such as sex or age)  more than another, even though the rules are formally neutral.  U.S.~labor law prohibits employers from using tests or selection procedures that are not ``job-related for the position in question and consistent with business necessity'' and that demonstrate adverse impact. 
While there is no single specific threshold or test to prove discrimination on the basis of adverse impact, a typical metric is the four-fifths rule. The rule  states that if the selection rate for a certain group is less than 80 percent of that of the group with the highest selection rate, there may be adverse impact on that group. 
The four-fifths rule does not apply to selection of military personnel, but similar principles hold given the military's commitment to equal opportunity.
The average gold-standard pass rates of the ACFT measured over six months of initial testing on over 14 thousand soldiers was 32 percent for women and 89 percent for men. The Army did not provide further disaggregation by age, nor  did the Army provide a breakout using gray or black standards, which are higher rates for certain career fields.%
\endnote{Gold-standard pass rates on the ACFT (from Army memo dated 09 July 2020 in response to Senator Gillibrand's request for information):\\
	\begin{tabularx}{1\linewidth}{lccccccc}\toprule
	 &FY19 & Oct & Nov & Dec & Jan & Feb & Average\\\midrule
	Female &21&37 & 29 & 33 & 34 & 40 & 32 \\
	Male  &  81& 91 & 89 & 91 & 90 & 90 & 89\\\midrule
	\end{tabularx}}
The impact ratio (32 divided by 89) is 36 percent, which is much lower than the 80 percent threshold needed to demonstrate adverse impact.

In assessing test fairness RAND researchers argue that \textQT{adverse impact alone does not indicate that a test is unfair to the group affected. A test could show adverse impact for women, but it could still be a fair and accurate predictor of their ability to do the job.} Instead, they cite predictive bias as a second consideration of fairness.  Predictive-validity bias refers to a test's accuracy in predicting the performance of a group of soldiers.  A test would be considered ``unbiased'' if it predicted performance equally well for all groups of soldiers.  If a test is a better predictor of one group than another, then the test is considered biased against the group with lower predictive validity. It is quite clear that the leg tuck test event is biased against women. Leg tucks, which have a 92-to-40 male-to-female gold standard pass rate, were shown to have no impact on predictive performance in the Fort Benning model. Even in the Fort Riley model leg tucks account for only four percent of predictive performance.  In other words, according to the predictive model, each leg tuck is equivalent to roughly two push-ups, a two-pound increase in deadlift, or five seconds of the two-mile run. Yet, in going from gold to gray standard on the ACFT, the Army has set each leg tuck equivalent to roughly ten push-ups, a twenty-pound increase in deadlift, or sixty seconds of the two-mile run.  The best strategy for a soldier to maximize his or her ACFT score, is by doing more leg tucks, a strategy that is most easily accomplished by male soldiers, yet one that has been shown to have little if any impact on the predicted performance on the \WTST.

For years, the military services have had different  standards for general fitness and for occupational requirements.
RAND researchers argue: \textQT{A fitness standard developed with the goal of improving overall health may determine that minimally acceptable fitness levels could be higher for younger, male personnel. In contrast, fitness standards developed with the goal of ensuring physical readiness to perform occupationally relevant, physically demanding tasks may use one standard for all personnel expected to perform those physically demanding tasks.}\autocite{RAND2017fit} In another study RAND researchers argue that military services traditionally have set two types of physical standards. General fitness standards, used to promote overall health status and physical fitness regardless of occupation, need not be gender neutral. The researchers state that ``it could be reasonable to continue to use gender-specific standards for physical fitness requirements for enlistment and continuation in service, while still requiring that occupation-specific physical standards be gender neutral, if the goals of the two standards are different.''  They explain that the goal of occupation-specific standards is ``to ensure that people are capable of performing a specific set of tasks required of everyone or considered critical to performance on the job,'' regardless of gender or age. Fitness standards, on the other hand, could be in place ``to maintain a culture of military discipline, bearing, and appearance; to keep health care costs to a minimum; to ensure personnel are not likely to be hampered by chronic illness; and to ensure that the personnel hired reflect the portion of the U.S. population who are at the peak of their health.'' Specifically, they state ``All of these goals can be achieved using screening tools that evaluate someone's overall medical health and fitness. However, in general, medical health and fitness measures used for these purposes are gender-normed and age-normed.''\autocite{RAND2018establishing1}
The Army itself has argued that the the ACFT is part of its broader Holistic Health and Fitness Program, whose stated goals are to 
improve individual Soldier readiness, transform the Army culture of fitness, reduce preventable injuries and attrition, enhance mental toughness and stamina, and contribute to increased unit readiness.\autocite{army2020fm} From the Army's own framework, the ACFT is best categorized as a general fitness assessment and as such ought to be gender- and age-normed.




\sparagraph{Recommendations}
It is  clear  under any scrutiny that the BSPRSS predictive model  is wrong. 
Recounting just one example---the model predicts a mean performance of 260 seconds on the \WTST\ using ACFT test events, when the true mean performance time was over 600 seconds. 
That neither  Army researchers nor  University of Iowa reviewers  caught such a gross error indicates that the model was not adequately scrutinized.
The 2015 NDAA requires that occupational standards accurately predict a soldier's performance on actual duties and that they are applied equitably.  Even when corrected for gross errors, it is unlikely from the Army's own study that leg tucks have any substantive predictive measure in the model.  Returning to George Box's aphorism about the wrongness and usefulness of models, the following are recommendations for Army researchers and leadership on making the BSPRSS model less wrong and more useful:%
\endnote{Also see \href{https://kylenovak29.s3.amazonaws.com/Thinking+Critically+about+Decision+Models.pdf}{``How to be Less Wrong and More Useful: Thinking Critically about Decision Models''}, Mar 1, 2020}
\begin{itemize}
\item 
Pause implementation of the ACFT as fitness test of record until the errors and concerns can be  addressed.
Reexamine the BSPRRS model and the training data to correct all gross errors and account for anomalies.
\item 
Establish ACFT success criteria based on success criteria on \WTST\ or other real-world tasks. Because success criteria were never identified on \WTST\ vignettes, there is no way set non-arbitrary ACFT standards.  For example, what is the maximum acceptable number of seconds allowed to extract and evacuate a casualty or to move over an obstacle? 
\item 
Be truthful about the model performance and limitations. Stating that the ACFT is over 80 percent predictive is bullshit.\autocite{bergstrom2020calling} 
\item 
Openly publish the model and data in machine-readable format to increase transparency and follow best practices in evidence-based policymaking.%
\endnote{\href{https://www.govinfo.gov/content/pkg/PLAW-115publ435/pdf/PLAW-115publ435.pdf}{Public Law 115 - 435 - Foundations for Evidence-Based Policymaking Act of 2018}}
Use exploratory visualizations to help identify patterns and anomalies in the data and to make the results accessible and compelling to broad audiences.
\item 
Use design of experiments in developing a operationally-relevant fitness standard to examine the interactions of multiple parameters. Perform factor and sensitivity analysis of the ACFT events and \WTST\ vignettes along with other variables such as height, weight, body mass, and gender to determine the relative importance of each variable and identify hidden biases.
\item 
Cross-validate the BSPRRS model using soldiers who are representative of all ages, genders, and military occupations.  Better yet, use a representative sample of soldiers in developing a new model. 
\item 
Think critically about model design and data collection. Continue working with external experts such as the University of Iowa Virtual Soldier Center in developing a scientifically rigorous standard. Include a diverse range of experts early in the design process to avoid group think. Red team the model to identify shortcomings.
\item 
Clearly delineate the function of the ACFT as either an operationally predictive, gender-neutral standard or as a component of the Holistic Health and Fitness Program meant to address injuries and general fitness.   
\end{itemize}


\printendnotes[aubcustomlist]          
\null\vfill
\setlength{\fboxsep}{0.2cm}
\fcolorbox{FontBlue}{white}{
\parbox{0.89\linewidth}{\footnotesize This report was written by Kyle A. Novak, Ph.D.,  a legislative fellow working in a personal office of the U.S.~Senate. The fellowship was supported through a grant by the Alfred P. Sloan Foundation and sponsored by the American Statistical Association and five other mathematical-statistical societies to improve evidence-based policymaking in the federal government. The views expressed in this publication are solely those of the author.\\[1\baselineskip]
Updated: November 19,~2020
\hfill \includegraphics[width=0.2\linewidth]{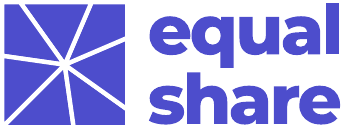}}}
\end{document}